# High performance construction materials fracture and high cycle fatigue assessment based on accelerated PF-CZM

*Jiaqi Li*[1], *Zhihua Xiong**[1], *Xuyao Liu*[2], *Hongyu Liu*[1]

*1. College of Water Resources and Architectural Engineering, Northwest A&F University, Yangling, China*

*2. College of Civil Engineering and Architecture, Zhejiang University, Hangzhou, China*

*Corresponding author: zh.xiong@nwsuaf.edu.cn*

**ABSTRACT:** Given the widespread application of high-performance materials in cyclic loaded infrastructure, a thorough understanding of the fracture and high-cycle fatigue behavior of high-performance materials is of critical importance. However, these behaviors predominantly rely on experimental methods, which are often costly, time-consuming and limited in generalizability. To address these issues, this paper proposes the constitution and high-cycle fatigue of Ultra-High Performance Concrete (UHPC) and High Strength Steel (HSS) within the Phase-Field Cohesive Zone Model (PF-CZM) framework. A generic stress-based failure criterion for UHPC is derived from the biaxial tensile test results to calculate crack driving force. Furthermore, a fatigue degradation function and an acceleration algorithm are integrated into the PF-CZM framework to enable efficient high-cycle fatigue simulations. In the acceleration algorithm, the envelope load is used to approximate the real cyclic load to avoid the simulation of each cycle, and the three-stage fatigue process is simulated through adjusting the cyclic increment adaptively. The proposed model successfully captures mixed-mode fracture and high-cycle fatigue behavior of UHPC and HSS, with validation provided through relevant experimental comparisons.

**Keywords:** high performance material; Phase-field model; accelerated algorithm; fatigue; high cycle; fracture.

## 1. Introduction

High performance materials, such as High Strength Steel (HSS) and Ultra-High Performance Concrete (UHPC) are integral to demanding applications in civil and mechanical engineering due to their superior mechanical properties [1-4]. In structures like wind turbines and bridges, they are routinely subjected to cyclic loading, where fracture and fatigue phenomena can lead to catastrophic failures [5-6]. However, modeling the fracture and fatigue behavior of these materials remains a significant challenge, especially for High Cycle Fatigue (HCF) scenarios involving millions of load cycles, which are often computationally prohibitive to simulate. For the above issues, Phase-Field Cohesive Zone Model (PF-CZM) is a promising solution.

Phase Field Models (PFM) originated from the linearly elastic variational principle, which is a powerful framework for simulating complex crack. It avoids pre-defining the crack path in Griffith's energy theory, more importantly it enables predictions of the crack nucleation, propagation and branching [7-8]. Based on PFM, the PF-CZM was introduced by Wu to better model both brittle and quasi-brittle fracture [9]. This is achieved by incorporating a parameterized crack geometry function and an energy degradation function, which respectively characterize the distribution of phase field crack and the strain energy dissipation during failure. Consequently, PF-CZM accurately predicts multiple softening curves and mitigates the mesh sensitivity associated with the length-scale parameter in traditional PFM, proving its utility in the fracture analysis of various materials [10-13].

To apply the PFM to fatigue analysis, two approaches are commonly employed: 1) introducing fatigue degradation functions related to historical variables to reduce fracture toughness [14-15], 2) adding an additional energy term to explain irreversible fatigue damage, *i.e.*, considering the fatigue mechanism occurring in the material as an additional contribution to the driving force [16-17]. In the analysis of high cycle fatigue, both of the above approaches are computationally intensive due to the presence of a large number of load cycles. To overcome this bottleneck, several fatigue acceleration algorithms for brittle fracture in PFM have been developed, including adaptive cyclic increment adjustment algorithm [18-20], cycle jump method [21-22] and adaptive meshing method [23-24]. These methods reduce computation time by adjusting the cyclic increment, jumping cycles and reducing the number of meshes, respectively.

Despite these advances, significant research gaps persist for specific high-performance materials. For fracture and fatigue of Ultra-High Performance Concrete (UHPC), experimental studies have explored the effects of aggregate size, steel fiber volume fractions etc. [25-27], leading to empirical fatigue life predictions [28]. However, numerical simulations for fatigue of UHPC are scarce, often limited to low-cycle fatigue [29] or are unable to reproduce the complete three-stage crack propagation process [30]. Similarly, for High Strength Steel (HSS), research has focused on experimental characterization of its ductile-to-brittle transition at low temperatures [31-33], with microstructural analyses explaining the underlying mechanisms [34-35]. The brittle fracture of large-scale screws made of HSS at low temperatures was also studied [36]. Yet, robust numerical models capable of predicting the low-temperature fracture behavior are notably absent.

In general, most of the research on the fracture and fatigue performance of typical high performance materials are based on experiments, which are often costly, time-consuming and limited in generalizability. This paper aims to bridge these gaps by developing a comprehensive and computationally efficient framework for modeling fracture and HCF in high-performance materials. We extend the PF-CZM to HCF analysis by integrating a fatigue degradation function with an acceleration algorithm. Specifically, for UHPC, we propose a new constitutive model and a generic stress-based failure criterion, calibrated against uniaxial and biaxial test data. Within this unified framework, mode-I and mixed-mode fracture simulation of UHPC and HSS at low temperature are achieved, and further investigation is conducted on the high cycle mode-I and mixed-mode fatigue of UHPC. The accuracy of the proposed model is verified through relevant tests.

## 2. Phase-field cohesive zone model

### 2.1 PF-CZM for fracture

Given a solid $\Omega \subset \mathbb{R}^n$ ($n$ = 1, 2, 3) with sharp crack, and its outer boundary and outer normal vector are denoted as $\partial\Omega \subset \mathbb{R}^{n-1}$ and $\mathbf{n}$, respectively. The body force $\boldsymbol{b}^*$ is distributed throughout the whole domain $\Omega$. The outer boundary $\partial\Omega$ is divided into two parts, $\partial\Omega_u$ and $\partial\Omega_t$, and given displacement boundary $\boldsymbol{u}^*(x)$ and force boundary $\boldsymbol{t}^*(x)$ are applied respectively, where $x$ is the spatial coordinate. The sharp crack is denoted as $S \subset \mathbb{R}^{n-1}$, with the normal vector denoted as $\mathbf{n}_S$.

In PFM, the sharp crack $S$ is regularized into a finite scale crack band $B \subseteq \Omega$, with its outer boundary denoted as $\partial$B and the outer normal vector denoted as $\mathbf{n}_B$. It should be noted that during the process of solid facture, the crack band $B$ is not pre-set or kept invariant, but continuously updated dynamically. Assuming small strain $\varepsilon(\mathbf{u})$, in order to quantify the energy stored and dissipated in the material, the work density function $W$ is divided into solid strain energy and crack surface energy, as written in Eq. (1) [37-38].

$$W(\mathbf{u},d)=\int_{\Omega}\Psi\left[\boldsymbol{\varepsilon}(\mathbf{u}),d\right]\mathrm{d}V+\int_{B}G_{\mathrm{f}}\gamma(d,\nabla d)\mathrm{d}V \quad (1)$$

Where $\gamma(d,\nabla d)$ is the density function of the crack surface and $d$ is the scalar PF parameter, which varies from 0 (undamaged material) to 1 (fully damaged material). $G_{\mathrm{f}}$ is the fracture energy of the material.

Using the PF parameter $d$ and its gradient $\nabla d$, the crack surface density function $\gamma(d, \nabla d)$ with a unit of $\mathrm{m}^{-1}$ is defined as:

$$\gamma(d,\nabla d)=\frac{1}{c_0}\left[\frac{1}{b}\alpha(d)+b|\nabla d|^2\right] \quad (2)$$

Where $b$ is a length scale regularizing the sharp crack, $c_0$ is the scaling factor which is expressed as:

$$c_0=4\int_0^1\sqrt{\alpha(\beta)}d\beta \quad (3)$$

where $\alpha(d)$ is the crack geometric function, which must satisfy the following conditions:

$$\alpha(0)=0,\quad \alpha(1)=1 \quad (4)$$

In PF-CZM, the dimensionless crack geometry function $\alpha(d)$ is defined as [9]:

$$\alpha(d)=\xi d+(1-\xi)d^2 \quad (5)$$

The shape of the crack geometry function under different values of $\xi$ is shown in Fig. 1 (a). It can be seen from the Fig. 1 (a) that the geometric functions of PF-CZM are different from traditional PFM AT1 and AT2, which use linear functions $\alpha(d) = d$ and quadratic functions $\alpha(d) = d^2$ to describe brittle fracture, respectively.

To describe the degradation of initial strain energy during the crack propagation, a monotonically decreasing energy degradation function which is dimensionless is defined as [37]:

$$\omega(d)=\frac{(1-d)^m}{(1-d)^m+a_1dP(d)} \quad (6)$$

where $a_1$ is a parameter inversely proportional to the length scale $b$. The smaller its value, the higher the accuracy of the regularization crack area (Eq. (2)), the smaller the range of influence of the crack on the surrounding solid, the closer the regularization crack is to the real crack, and the steeper the energy degradation function $\omega(d)$, as shown in Fig. 1 (b). $P(d) = 1+a_2d+a_3d^2$, the values of $a_2$, $a_3$, and $m$ can be determined by the cohesive law that controls the softening behavior of the material, which represents the stress $\sigma$ as a function of the crack opening $w$.

In PF-CZM, assuming a bar specimen used for fatigue crack propagation tests, which is sufficiently long to ensure that crack evolution is not affected by boundary effects. Displacement loads in opposite directions are applied to both ends of the specimen, and distributed body forces are neglected. For simplicity, it is assumed that the crack initiates at the symmetric point $x = 0$. Then for a given crack geometry function and energy degradation function, the stress with a unit of MPa and apparent displacement jump with a unit of mm across the crack band is defined as [9]:

$$\sigma(d^*) = f_t\sqrt{\frac{[\xi+(1-\xi)d^*](1-d^*)^m}{\xi P(d^*)}} \quad (7)$$

$$w(d^*) = \frac{4G_f\sqrt{\xi}}{c_0 f_t}\int_0^{d^*}[\frac{P(d^*)}{(1-d^*)^m}\cdot\frac{\xi+(1-\xi)\beta}{\xi+(1-\xi)d^*}-\frac{P(\beta)}{(1-\beta)^m}]^{-\frac{1}{2}}\frac{\sqrt{\beta}\cdot P(\beta)}{(1-\beta)^m}d\beta \quad (8)$$

Where the failure strength $f_t$ is given as:

$$f_t = \sqrt{\frac{2E_0G_f}{c_0\cdot b}\cdot\frac{\xi}{a_1}} \quad (9)$$

Where $E_0$ is the Young's modulus of the material and the value of the parameter $P(d^*)$ is determined by the maximum damage $d^*$ attained at the symmetric point $x = 0$.

Eq (7) and (8) provide a general softening law parameterized by the maximum damage $d^*$. The initial slope $k_0$ and ultimate crack opening $w_c$ of the softening curve could be expressed as:

$$k_0 = -\frac{c_0}{4\pi}\cdot\frac{f_t^2}{G_f}\cdot\frac{[\xi(a_2+m+1)-1]^{3/2}}{\xi^2} \quad (10)$$

$$w_c = \frac{2\pi G_f}{c_0 f_t}\sqrt{\xi P(1)}\lim_{d^*\to 1}(1-d^*)^{1-m/2} \quad (11)$$

The parameters $a_1$, $a_2$, $a_3$ are expressed as [9]:

$$a_1 = \frac{2E_0G_f}{f_t^2}\cdot\frac{\xi}{c_0\cdot b} \quad (12)$$

$$a_2 = 2\beta_k^{2/3}-(m+\frac{1}{2})\ldots-\frac{1}{2} \quad (13)$$

$$a_3 = \begin{cases} 0, & m>2 \\ \frac{1}{2}\beta_w^2-(1+a_2), & m=2 \end{cases} \quad (14)$$

Where $\beta_k$ and $\beta_w$ are defined as:

$$\beta_k = \frac{2k_0}{-f_t^2/G_f}\ldots 1 \quad (15)$$

$$\beta_w = \frac{w_c}{2G_f/f_t} \quad (16)$$

It is found that the values of $a_2$ and $a_3$ are independent of the length scale $b$. In addition, as long as the values of $\xi$, $b$, and $m$ are determined, the values of $a_1$, $a_2$, and $a_3$ can be determined based on the softening curve of the target material. Fig. 1 (c) describes the commonly used softening curves for quasi brittle materials.

The solid strain energy density is expressed as:

$$\psi[\varepsilon(\mathbf{u}),\omega(d)] = \omega(d)\psi_0(\varepsilon) \quad (17)$$

Where $\Psi_0(\varepsilon)$ represents the initial strain energy density, which is written as:

$$\psi_0(\varepsilon) = \frac{1}{2}\varepsilon:\mathbb{C}:\varepsilon = \frac{1}{2}\bar{\sigma}:\mathrm{S}:\bar{\sigma} \quad (18)$$

Where $\mathbb{C}$ is elastic stiffness, $\mathrm{S}$ is compliance, $\bar{\boldsymbol{\sigma}}$ is effective stress tensor, which could be expressed as:

$$\bar{\sigma} = \mathbb{C}:\varepsilon \quad (19)$$

The apparent stress tensor is written as:

$$\boldsymbol{\sigma} = \frac{\partial\psi[\varepsilon(\mathbf{u}),\omega(d)]}{\partial\varepsilon} = \omega(d)\bar{\boldsymbol{\sigma}} = \omega(d)\mathbb{C}:\varepsilon \quad (20)$$

The conjugate energy release rate could be expressed as [39-40]:

$$Y = -\frac{\partial\psi[\varepsilon(\mathbf{u}),\omega(d)]}{\partial d} = -\omega'(d)\mathrm{Y} \quad (21)$$

Where $\mathrm{Y}$ represents the initial elastic strain energy, which is expressed as:

$$\mathrm{Y} = \psi_0(\varepsilon) \quad (22)$$

The coupled field equation of PF-CZM could be written as:

$$\nabla\cdot\boldsymbol{\sigma} + \mathbf{b}^* = \mathbf{0} \quad \text{in} \quad \Omega \quad (23)$$

$$\omega'(d)\mathrm{Y} + G_f\delta_d\gamma \geq 0 \quad \text{in} \quad B \quad (24)$$

Where the variation of the crack surface density function $\delta_d\gamma$ is expressed as:

$$\delta_d\gamma = \partial_d\gamma - \nabla\cdot\partial_{\nabla d}\gamma = \frac{1}{c_0}\left[\frac{1}{b}\alpha'(d) - 2b\Delta d\right] \quad (25)$$

Where $\Delta d$ represents the Laplacian operator $\Delta d = \nabla\cdot\nabla d$. Substituting Eq. (25) into Eq. (24) and considering the apparent stress tensor, the coupled field equation could be rewritten as [38]:

$$\nabla\cdot\left[\omega(d)\bar{\sigma}(\mathbf{u})\right] + \mathbf{b}^* = \mathbf{0} \quad (26)$$

$$\frac{\mathrm{G_f}}{c_0}\left[\frac{2}{l}(1-d) - 2b\Delta\mathrm{d}\right] = -\omega'(d)\mathrm{Y}(\mathbf{u}) \quad \text{for} \quad \dot{d} > 0 \quad (27)$$

The Neumann boundary conditions is expressed as:

$$\begin{cases} \bar{\boldsymbol{\sigma}} \cdot \boldsymbol{n} = \mathbf{t}^* & \text{on} \quad \partial \Omega_{\mathrm{t}} \\ \nabla d \cdot \mathbf{n}_B = 0 & \text{on} \quad \partial B \end{cases} \quad (28)$$

The stress-strain relationship and conjugate energy release rate clearly indicate that the material exhibits symmetry in tension and compression. Therefore, the material will not only undergo tensile fracture, but also compressive fracture. To prevent compression fracture of the material, the crack driving force is modified as follows:

$$\mathrm{Y} = \frac{1}{2E_0} \langle \bar{\boldsymbol{\sigma}}_{\mathrm{eq}} \rangle^2 \quad (29)$$

Where for mode I fracture, according to the Rankine criterion, the equivalent stress $\bar{\boldsymbol{\sigma}}_{\mathrm{eq}}$ could be written as:

$$\bar{\boldsymbol{\sigma}}_{\mathrm{eq}} = \langle \bar{\boldsymbol{\sigma}}_1 \rangle \quad (30)$$

For mixed mode fracture, according to the generic stress-based failure criterion, the equivalent stress could be expressed as [41]:

$$\bar{\boldsymbol{\sigma}}_{\mathrm{eq}} = \frac{\rho_s - 1}{2\rho_s} \bar{I}_1 + \frac{1}{2\rho_s} \frac{1}{\sqrt{2 - A_0}} \sqrt{\left[ (2 - A_0)\rho_s^2 + 2(3A_0 - 2)\rho_s + 2 - A_0 \right] \bar{I}_1^2 + 24\rho_s \bar{J}_2} \quad (31)$$

Where $\rho_s = f_c / f_t$ is the ratio of uniaxial compressive strength to uniaxial tensile strength, $\bar{\boldsymbol{\sigma}}_1$ is the first principal stress, $\langle x \rangle = \max\{0, x\}$ is the Macaulay bracket, $\bar{I}_1 = \mathbf{tr}\bar{\boldsymbol{\sigma}}$ is the first invariant, and $\bar{J}_2 = \frac{1}{2}\bar{\boldsymbol{\sigma}} : \bar{\boldsymbol{\sigma}} - \frac{1}{6}\bar{I}_1^{\,2}$ is the second invariant. The criterion is formulated in terms of the three principal stresses, which implicitly captures the effects of shear stress without its explicit inclusion.

In plane stress condition, the above failure criteria can be divided into the following four situations based on the different values of parameter $A_0$:

$$\begin{array}{ll} A_0 < \frac{1}{2} & \text{Elliptical function} \\ A_0 = \frac{1}{2} & \text{Parabolic function} \\ \frac{1}{2} < A_0 < 2(\rho_s^2 - \rho_s + 1)/(\rho_s + 1)^2 & \text{Hyperbolic function} \\ A_0 = 2(\rho_s^2 - \rho_s + 1)/(\rho_s + 1)^2 & \text{Two straight lines} \end{array} \quad (32)$$

The biaxial strength envelopes of the four situations are shown in Fig. 1 (d).

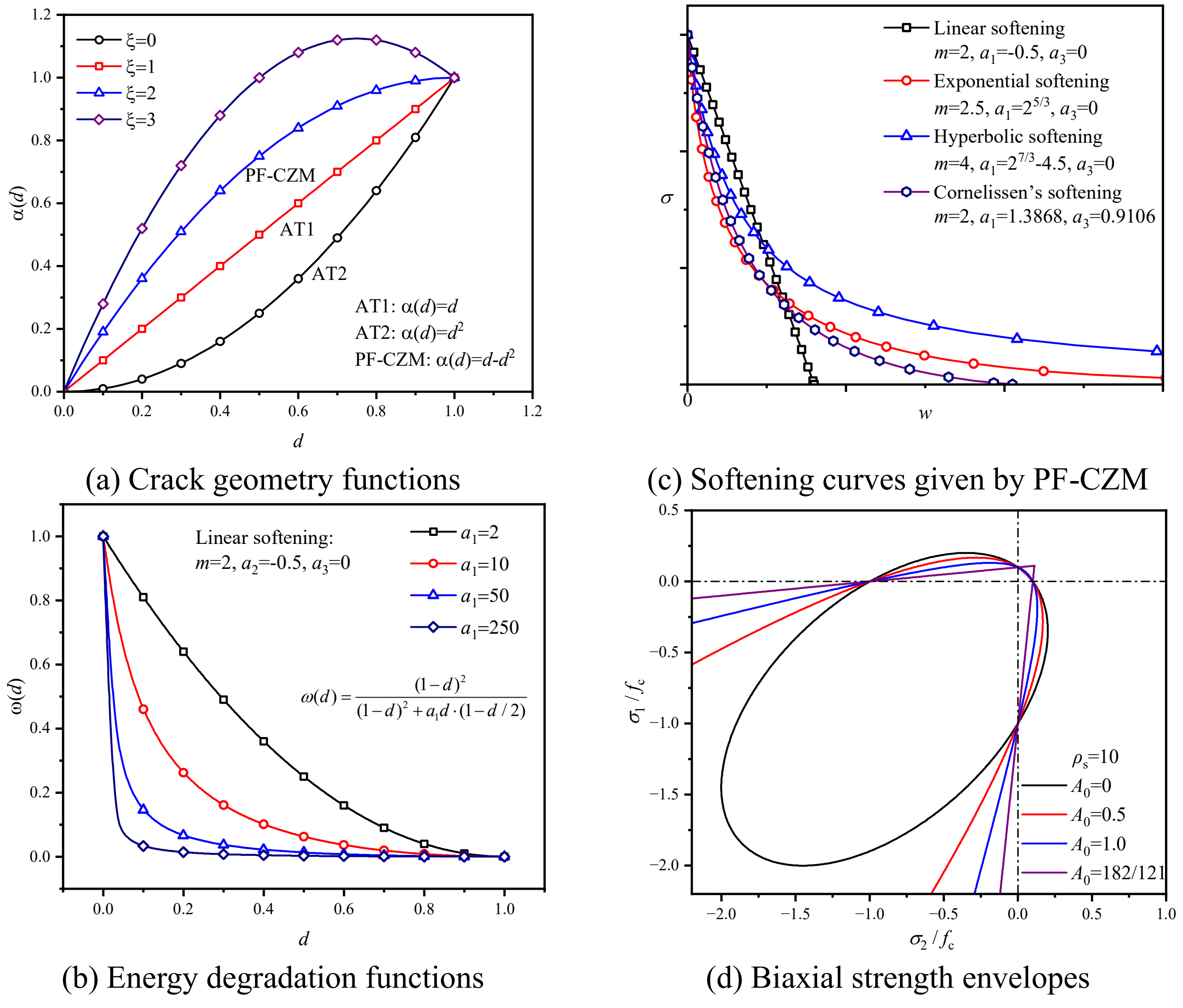

(a) Crack geometry functions

(c) Softening curves given by PF-CZM

(b) Energy degradation functions

(d) Biaxial strength envelopes

Fig. 1. Key functions of PF-CZM

Considering that material damage is irreversible, a historical field variable $\mathcal{H}$ is introduced to satisfy the following conditions:

$$d_{t+\Delta t} \geq d_t \quad (33)$$

Where $d_{t+\Delta t}$ represents the PF parameter in the current time increment, and $d_{\mathrm{t}}$ represents the PF parameter in the previous time increment. This historical field variable must satisfy the Kuhn-Tucker conditions:

$$\mathrm{Y}-\mathrm{H} \leq 0, \quad \dot{\mathrm{H}} \geq 0, \quad \dot{\mathrm{H}} \cdot (\mathrm{Y}-\mathrm{H}) = 0 \quad (34)$$

Therefore, the history field variable at the current time step $t$, within the total time $\tau$, can be expressed as:

$$\mathrm{H} = \max_{t\in[0,\tau]} \mathrm{Y}\left(t\right) \quad (35)$$

The minimum value of crack driving force is defined as [42]:

$$\mathrm{H}_{\min}=\frac{f_{\mathrm{t}}^{2}}{2E_{0}} \quad (36)$$

## 2.2 PF-CZM for fatigue

This part mainly adopts the authors' previous work in [46]. By directly incorporating the fatigue degradation function into the Griffith energy balance equation to consider the fatigue damage of materials [43], the concept of state dependent fracture toughness is adopted. The second coupled field equation (Eq. (27)) is modified to:

$$\frac{f\left(\bar{\alpha}(\mathrm{t})\right)\mathrm{G}_{\mathrm{f}}}{c_{0}}\left[\frac{2}{b}(1-d)-2b\Delta\mathrm{d}\right]=-\omega'(d)\mathrm{H} \quad \text{for} \quad \dot{d}>0 \quad (37)$$

Where $t$ is the pseudo-time, $\bar{\alpha}(t)$ is the cumulative historical variable. The function $f(\bar{\alpha}(t))$ is the fatigue degradation function, and its properties are shown in Eq. (38):

$$\begin{aligned}
&f\left(\bar{\alpha}(t)\le\alpha_{\mathrm{T}}\right)=1\\
&f\left(\bar{\alpha}(t)>\alpha_{T}\right)\in[0,1]\\
&f'\left(\bar{\alpha}(t)\right)\le 0 \quad \text{in} \quad 0\le f\left(\bar{\alpha}(t)\right)<1
\end{aligned} \quad (38)$$

The fatigue degradation function which is dimensionless is as follows [15]:

$$f\left(\bar{\alpha}(t)\right)=\begin{cases}1 & \text{if} \quad \bar{\alpha}(t)\le\alpha_{\mathrm{T}}\\ \left(\dfrac{2\alpha_{\mathrm{T}}}{\bar{\alpha}(t)+\alpha_{\mathrm{T}}}\right)^{2} & \text{if} \quad \bar{\alpha}(t)>\alpha_{\mathrm{T}}\end{cases} \quad (39)$$

$\alpha_{\mathrm{T}}$ is the threshold that controls the onset of fatigue damage, which is defined as:

$$\alpha_{\mathrm{T}}=\frac{G_{\mathrm{f}}}{k_{\mathrm{f}}\cdot l} \quad (40)$$

Where $k_{\mathrm{f}}$ is the fatigue damage accumulation parameter, which controls the fatigue life of the material by regulating the fatigue damage accumulation rate. Fig. 2 shows the fatigue degradation function under different values of the parameter $k_{\mathrm{f}}$.

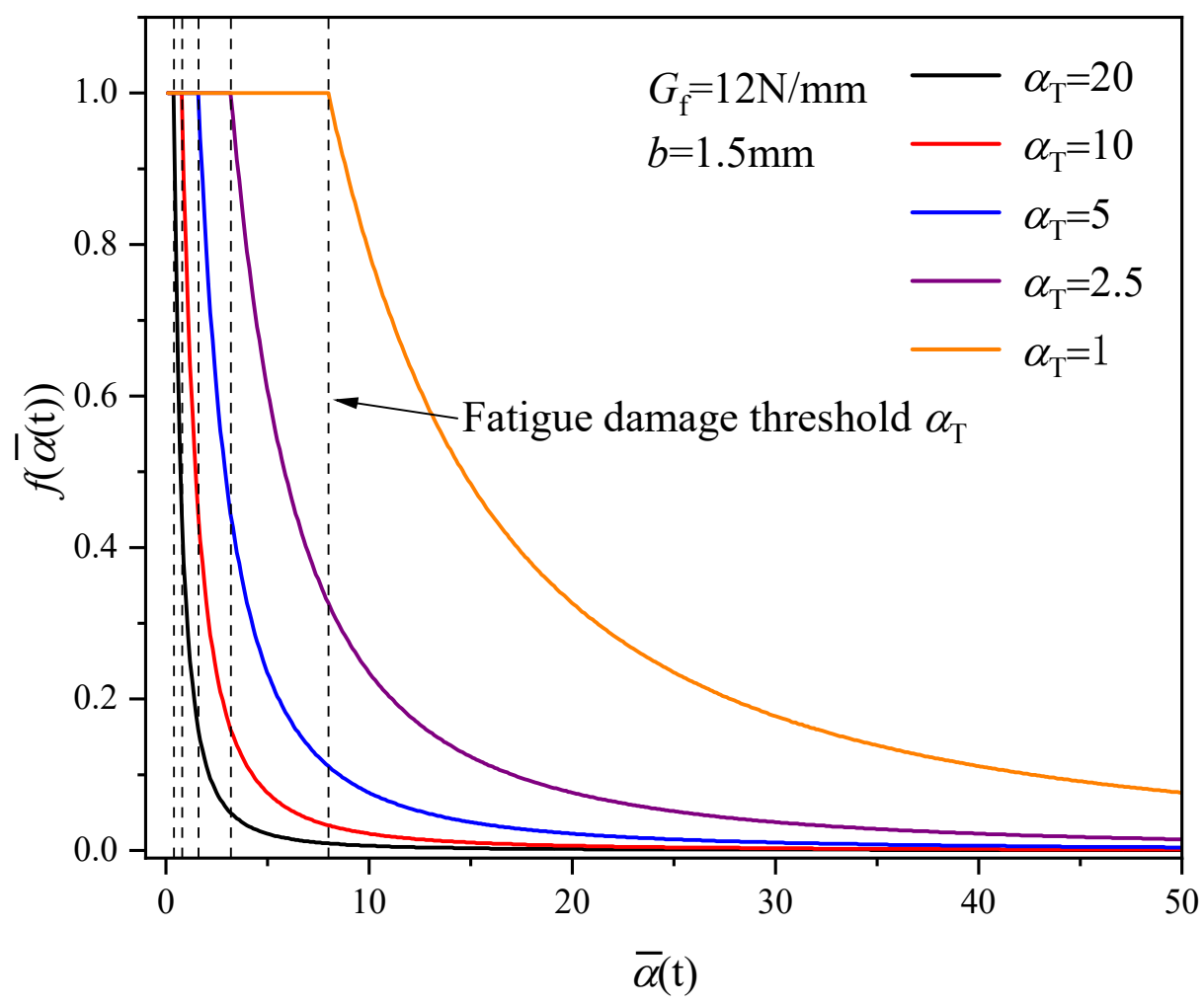

Fig. 2. Fatigue degradation function

According to the fatigue hypothesis [15,44], the cumulative historical variable $\bar{\alpha}(t)$ can be defined as:

$$\bar{\alpha}(t)=\begin{cases}\int_0^t\left|\dot{\alpha}\right|\mathrm{d}t & \text{if} \quad \alpha\dot{\alpha}\geq 0(\text{loading} \quad \text{stage}) \\ 0 & \text{if} \quad \alpha\dot{\alpha}<0(\text{unloading} \quad \text{stage})\end{cases} \quad (41)$$

Where $\alpha$ is the cumulative variable, according to [45], the formula is $\alpha(t) = [1\text{-}d(t)]^2\Psi_0(t)$.

## 2.3 Accelerated algorithm

An accelerated algorithm detailed in our previous study is employed here. It has been shown that throughout the entire fatigue life, the fatigue creep curves obtained by the accelerated algorithm and cycle-by-cycle simulation agree well, and the crack propagation processes are basically consistent. This validation confirms the algorithm's effectiveness while significantly reducing computational cost; for instance, the algorithm achieved a 97.62% reduction in computational time in a previous numerical example [46]. The envelope load is employed to approximate the cyclic load via a two-step fatigue simulation as illustrated in Fig. 3. The core of this approach is to link the cyclic increment d$N$ which is dimensionless with the damage increment d$D$ during each iteration. Unlike previous works [18-20], an analogy is drawn between the cumulative historical variable $\bar{\alpha}(t)$ and the damage variable $D$, where the cumulative historical variable $\bar{\alpha}(t)$ is defined as the damage variable D and $\alpha_t$ is labelled as the damage threshold $D_c$. The expressions for cyclic increment d$N$ are then given in Eq. (42-43):

$$\mathrm{d}N=n_D\left(\frac{\hat{\sigma}_n(1-L)}{f(L)}\right)^{-k_s}\mathrm{d}\bar{\alpha}(t) \quad (42)$$

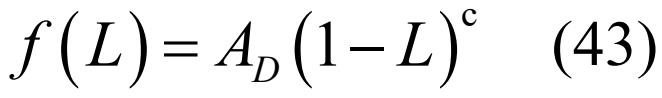

$$f(L) = A_D(1-L)^c \quad (43)$$

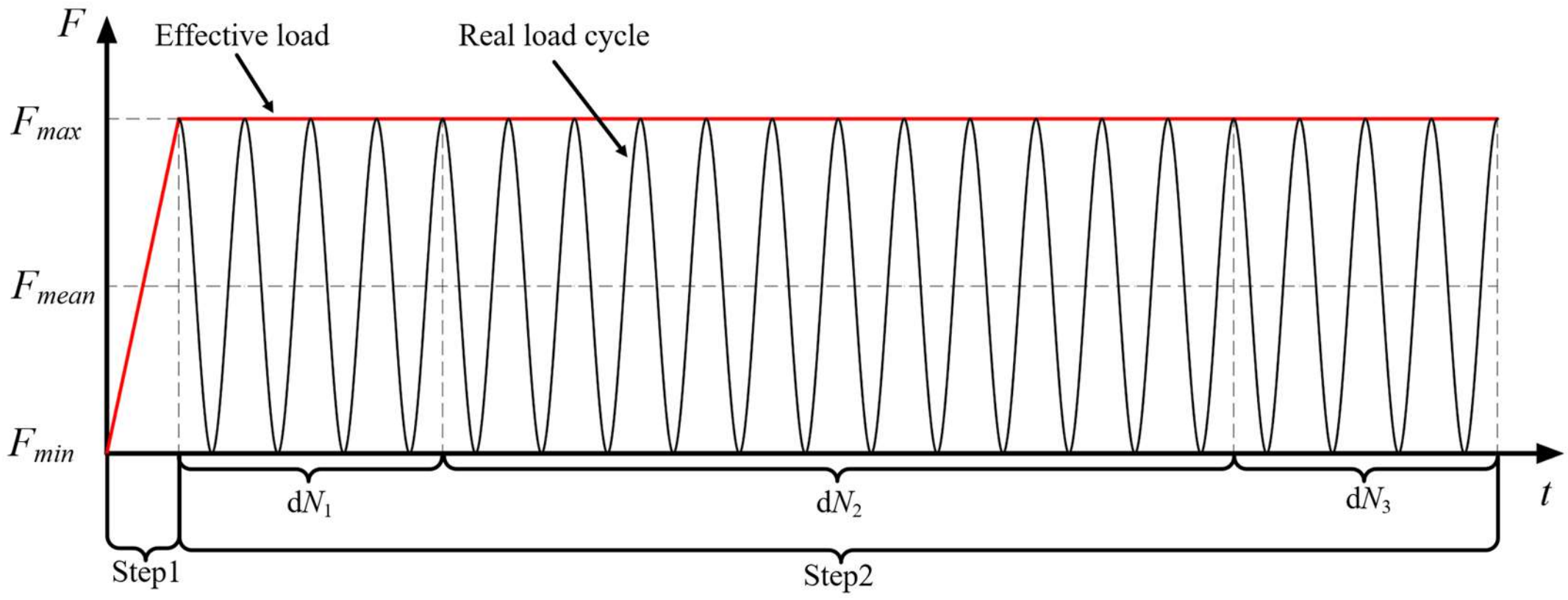

Fig. 3. Schematic diagram of two-step fatigue crack propagation simulation scheme

Where $f$(L) is the threshold, and $L$ is the ratio of the average load to the maximum load. The parameter $c$ is a constant and can be considered as the average stress effect. $\hat{\sigma}_n$ is the driving force and is taken as the maximum principal stress. The parameters $A_D$, $n_D$, and $k_s$ are obtained from the experimental Wöhler curve ($S$-$N$ curve). These parameters enable PF-CZM to directly combine experimental data, achieving intuitive simulation of complex fatigue crack evolution. The parameters related to the accelerated algorithm are summarized in Table 1 [46].

Table 1. Model parameters [46]

| Model parameters | Value |
|---|---|
| Slope Factor $k_s$ | 0.43 |
| Mean stress parameter $c$ | 0.5 |
| Knee point cycle number $n_D$ | 5000000 |
| Control parameters $\bar{\alpha}_2(t)$ | 0.01 |
| Control parameters $\bar{\alpha}_3(t)$ | 0.05 |

Since the cyclic increment d$N$ determines the simulation speed, we employ an adaptive cycle-increment adjustment algorithm. The fatigue simulation is partitioned into three stages—each with a different d$N$—based on the evolution of the damage variable $\bar{\alpha}(t)$.

1) Pre-threshold stage: the fatigue damage is below the damage threshold $\alpha_T$, d$N$ should be as large as possible.
2) Threshold stage: the fatigue damage approaches $\alpha_T$, the selection of d$N$ should ensure that the damage increment is sufficiently small.
3) Post-threshold stage: the fatigue damage exceeds $\alpha_T$, adaptive d$N$ is used to control

the maximum damage increment max[d$\bar{\alpha}(t)$] between the control parameters $\bar{\alpha}_2(t)$ and $\bar{\alpha}_3(t)$ to maintain a moderate fatigue damage rate.

The three-stage adaptive cycle increment adjustment algorithm is demonstrated in Fig. 4 [46]. The accelerated algorithm is shown in Algorithm 1 [46].

**Algorithm 1**: three-stage adaptive cycle increment adjustment algorithm [46]

```
Input: input parameter: αT, ᾱ1(t), ᾱ2(t)
/* damage threshold: αT */
/* regulation parameter at stage II: ᾱ1(t)*/
/* regulation parameter at stage III: ᾱ2(t)*/
/* regulation parameter at stage III: ᾱ3(t)*/
static is TRUE;
crack is FALSE;
while simulation do
if max[dᾱn+1(t)] + max[dᾱn(t)] <αT and static is TRUE
then
/* stage I */
dN = 2dN;
end
else if max[dᾱn+1(t)] + max[dᾱn(t)] >αT and
crack is FALSE then
/* stage II */
static is FALSE;
if max[dᾱn(t)] > ᾱ1(t) then
dN = 0.1dN;
restart;
end
else if max[dᾱn(t)] < ᾱ1(t)then
/* fatigue damage begins now */
crack is TRUE;
end
end
if crack is TRUE then
/* stage III */
if max[dᾱn(t)] < ᾱ2(t) then
dN = dN·ᾱ3(t) / ᾱ2(t);
restart;
end
else if max[dᾱn(t)] > ᾱ3(t) then
dN = dN·ᾱ3(t) / ᾱ2(t);
```

```
restart;
end
end
end
```

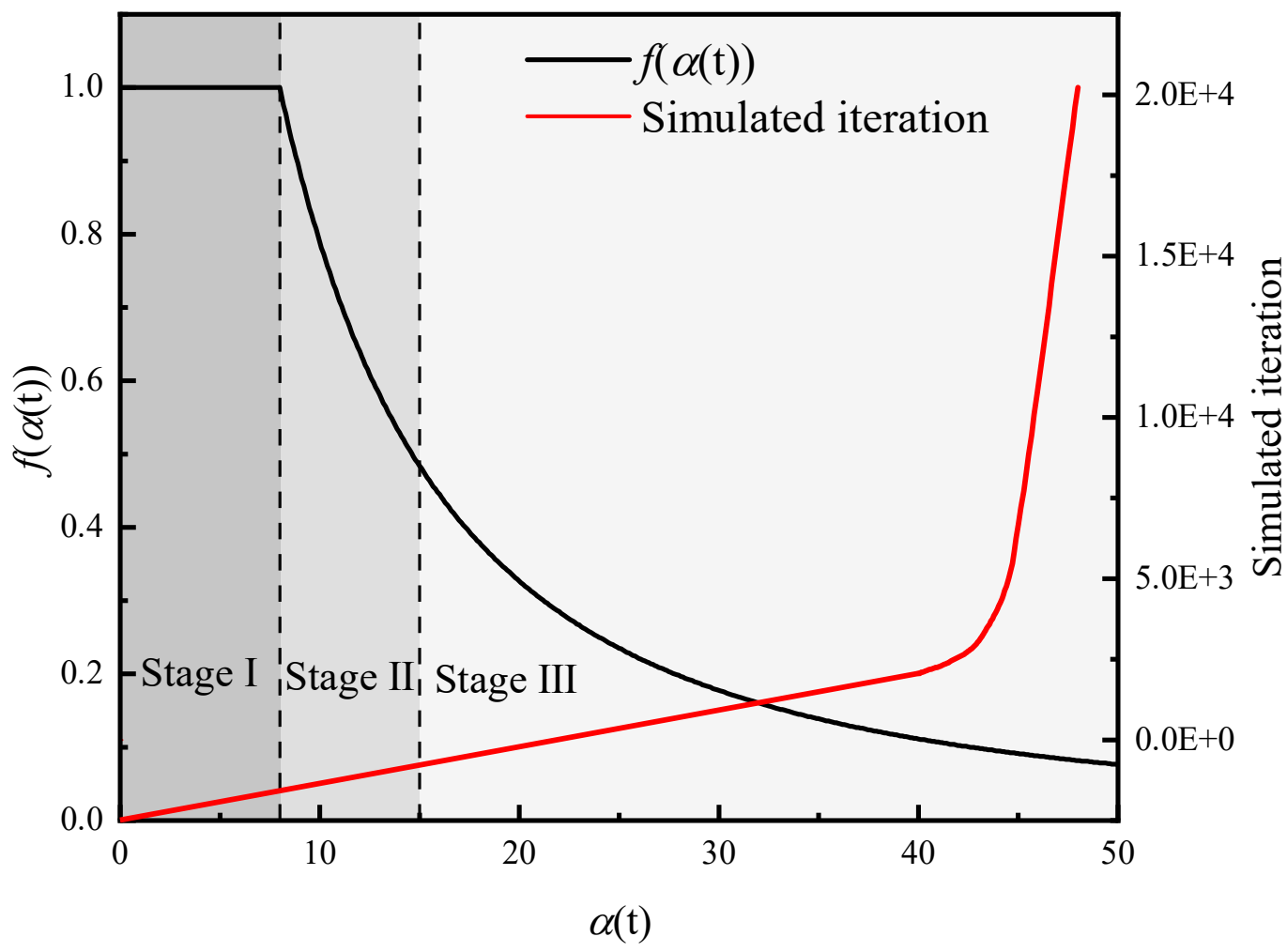

Fig. 4. Schematic illustration of the three-stage adaptive cycle increment adjustment algorithm [46]

## 2.4 Numerical implementation

PF-CZM described in Section 2.1 and 2.2 and accelerated algorithm described in Section 2.3 are implemented in ABAQUS using User Material (UMAT) and Heat Flux (HETVAL) subroutine. This implementation utilizes the analogy between heat transfer law and PF balance equation, using the temperature field $T$ instead of the PF parameter $d$, and making it also vary between 0 and 1 [42,47-48]. For example, for a solid with thermal conductivity $k$, specific heat capacity $c_\rho$, and density $\rho$, the temperature field $T$ over time $t$ under the action of an internal heat source $r$ is given by the following equilibrium law:

$$k\nabla^2 T - \rho c_P \frac{\partial T}{\partial t} = -r \qquad (44)$$

Under steady conditions, $\partial T/\partial t = 0$, the above law is simplified as:

$$k\nabla^2 T = -r \qquad (45)$$

Similarly, the PF balance equation could be written as:

$$\Delta d = \nabla^2 d = \frac{c_0}{2bf\left(\bar{\alpha}\right)G_\mathrm{f}}\omega'\left(d\right)\mathrm{H} + \frac{1}{b^2}\left(1-d\right) \qquad (46)$$

Assuming the unit thermal conductivity $k = 1$, the heat flux generated by heat source $r$ is defined as:

$$r=-\frac{c_0}{2bf\left(\bar{\alpha}\right)G_{\mathrm{f}}}\omega'\left(d\right)\mathrm{H}-\frac{1}{b^2}\left(1-d\right) \quad (47)$$

Thus, the rate of change of heat flux relative to temperature is expressed as:

$$\frac{\partial r}{\partial d}=-\frac{c_0}{2bf\left(\bar{\alpha}\right)G_{\mathrm{f}}}\omega''\left(d\right)\mathrm{H}+\frac{1}{b^2} \quad (48)$$

In the numerical implementation of PF-CZM, the PF variable $d$ is treated as the temperature field $T$. Consequently, the Laplacian operator acting on $d$ corresponds to the temperature Laplacian, and the right-hand side of the PF balance equation is mapped to the internal heat source term $r$. While each equation is dimensionally consistent on its own (W/m$^3$ for heat conduction and 1/m$^2$ for the PF equation), the two systems do not share the same physical units. The solver operates on the identical structure of the governing equations, not on their physical interpretation.

By using the CPE4T temperature-displacement coupling element in ABAQUS, this implementation can obtain integration point level results. At the integration point level in ABAQUS, the subroutine obtains strain tensor and PF parameters from the interpolation solved by nodes of specific elements. Firstly, in the UMAT subroutine, the stress tensor $\boldsymbol{\sigma}$ and material stiffness tensor $\mathbb{C}$ are calculated through the strain tensor. Then these tensors are degraded based on the current PF parameters. Finally, the cumulative history variable $\bar{\alpha}(t)$ and fatigue degradation function $f(\bar{\alpha}(t))$ are calculated. The crack driving force is stored in solution-dependent state variables (SDVs), thereby executing irreversibility and fatigue degradation function. In the HETVAL subroutine, the current values of the history field $\mathcal{H}$ and the fatigue degradation function $f(\bar{\alpha}(t))$ are converted based on the updated SDVs, and the internal heat flux $r$ (Eq. (47)) and its derivative with respect to temperature (Eq. (48)) are defined considering the fatigue degradation of fracture energy. The above process is repeated at each integration point. Then the element stiffness matrix and residual are constructed, thus the global equation system is built.

# 3 UHPC fracture and high cycle fatigue modeling

## 3.1 PF-CZM for UHPC

### 3.1.1 Crack geometry function and energy degradation function

In PF-CZM, the crack geometry function and energy degradation function determine its analysis results. Therefore, in order to study the fracture behavior of UHPC using PF-CZM, it is necessary to determine the applicable crack geometry function and energy degradation function firstly, so as to reasonably reflect the strain softening stage during the fracture process. $\xi = 2$ is recommended in reference [9], the crack geometry function is written in Eq. (49).

$$\alpha(d) = 2d - d^2 \quad (49)$$

To determine the energy degradation function of UHPC, it is necessary to determine the values of key parameters $a_1$, $a_2$, and $a_3$. According to Eq. (12)-(14), the value of $a_1$ is determined based on the relevant parameters input in PF-CZM, and $a_2$ and $a_3$ is determined by the initial slope and the maximum crack opening of the UHPC softening curve, which need to be calibrated before analysis. Saqif et al. [49] conducted uniaxial tensile tests on UHPC specimens using different types of steel fibers and steel fiber lengths, and obtained the parameterized formula of the UHPC softening curve as follows:

$$f = f_{\mathrm{t}}\left[1+\left(k_1\frac{w-w_o}{w_c-w_o}\right)^3\right]e^{-k_2\frac{w-w_o}{w_c-w_o}} - f_{\mathrm{t}}\frac{w-w_o}{w_c-w_o}(1+k_1^3)e^{-k_2} \quad (50)$$

Where $f_{\mathrm{t}}$ is the uniaxial tensile strength, $w$ is the crack opening, $w_{\mathrm{c}}$ is the ultimate crack opening, and $k_1$ and $k_2$ are parameters fitted based on experimental data. The values of parameters for different specimens in the formula are shown in Table 2.

Table 2. Parameters of UHPC softening curves [49]

| Specimen | $k_1$ | $k_2$ | $w_{\mathrm{c}}$/(mm) | $f_{\mathrm{t}}$/(MPa) | Fracture energy $G_{\mathrm{f}}$/(N/mm) |
|---|---|---|---|---|---|
| S1-13 | 2 | 3.2 | 6.1 | 11.03 | 21.97 |
| S2-13 | 1.9 | 4.7 | 6.3 | 10.55 | 19.92 |
| S2-19 | 1.3 | 4.1 | 10.4 | 10.74 | 22.55 |
| H-13 | 1.2 | 2.3 | 6.6 | 11.93 | 22.37 |

The initial slope $k_0$ of the UHPC softening curve can be obtained from Eq. (50) as follows:

$$k_0 = f'(0) = -\frac{f_{\mathrm{t}}}{w_c - w_0}\left[k_2 + (1+k_1^3)e^{-k_2}\right] \quad (51)$$

According to the initial slope and the maximum crack opening of the UHPC softening curve, the values of parameters $k_0$, $\beta_{\mathrm{k}}$, $\beta_{\mathrm{w}}$, $a_2$, $a_3$ are calculated and listed in Table 3.

Table 3. Parameters related to fatigue degradation function

| Specimen | $k_0$ | $\beta_{\mathrm{k}}$ | $\beta_{\mathrm{w}}$ | $a_2$ | $a_3$ |
|---|---|---|---|---|---|
| S1-13 | -6.449 | 2.329 | 1.531 | 1.014 | -0.842 |
| S2-13 | -9.541 | 5.018 | 1.733 | 3.362 | -2.859 |
| S2-19 | -4.288 | 1.676 | 2.476 | 0.322 | 1.743 |
| H-13 | -4.651 | 1.462 | 1.759 | 0.076 | 0.471 |

After determining the values of $a_2$ and $a_3$, the UHPC softening curves given by PF-CZM are obtained through Eq. (7)-(8), as shown in Fig. 5.

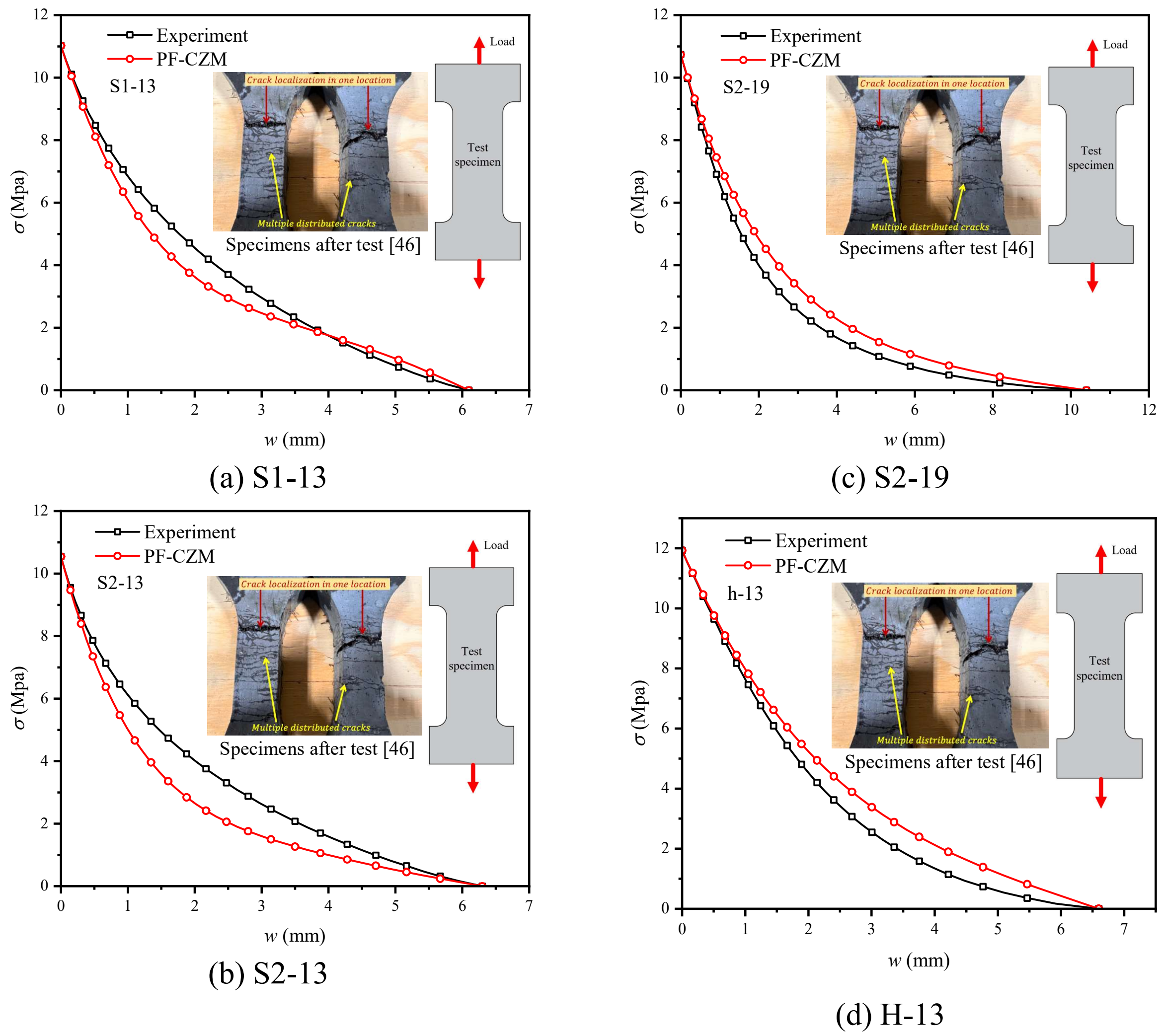

(a) S1-13 (c) S2-19 (b) S2-13 (d) H-13

Fig. 5. UHPC softening curves given by PF-CZM and experiments

3.1.2 Crack driving force

The calculation method of crack driving force is related to the fracture mode of the material according to Eq. (29)-(31). For mode-I fracture, the crack driving force is calculated based on the first principal stress (Eq. (30)), while for mixed-mode fracture, it follows the generic stress-based failure criterion (Eq. (31)) [41]. In Eq. (30), $\rho_s$ denotes the ratio of uniaxial compressive strength to uniaxial tensile strength, and $A_0$ is a parameter fitted based on experimental data, which needs to be calibrated before analysis. Liu et al. [50] conducted biaxial tensile and compression tests on a large number of UHPC specimens, and the experimental data are summarized in Table 4.

Table 4. Biaxial experimental data of UHPC [50]

| Specimen | $f_c$ (MPa) | $f_t$ (MPa) | $\rho_s = f_c/f_t$ | $\sigma_1$ (MPa) | $\sigma_2$ (MPa) |
|---|---|---|---|---|---|
| UP-CC-1 | 125.98 | 10.65 | 11.83 | -117.01 | -117.06 |
| UP-CC-2 | 127.33 | 10.65 | 11.96 | -152.97 | -76.43 |
| UP-CC-5 | 128.99 | 10.65 | 12.11 | -139.61 | -27.87 |
| UP-TT-1 | 130.02 | 10.65 | 12.21 | 8.40 | 8.66 |
| UP-TT-2 | 128.42 | 10.65 | 12.06 | 5.94 | 11.58 |
| US-TC-4.0 | 127.09 | 10.24 | 12.41 | 4.00 | -98.28 |
| US-TC-4.5 | 122.11 | 10.24 | 11.92 | 4.36 | -74.46 |
| US-TC-5.0 | 132.64 | 10.24 | 12.95 | 4.41 | -75.70 |
| US-TC-5.5 | 142.61 | 10.24 | 13.93 | 6.14 | -80.10 |
| US-TC-6.0 | 141.56 | 10.24 | 13.82 | 6.22 | -70.10 |
| UP-TC-1 | 127.44 | 10.71 | 11.90 | -9.23 | 9.16 |
| UP-TC-8 | 127.78 | 10.71 | 11.93 | -77.16 | 9.58 |
| UP-TC-10 | 128.44 | 10.71 | 11.99 | -76.07 | 7.59 |
| UP-TC-12 | 126.43 | 10.71 | 11.80 | -76.27 | 6.34 |
| Average | 129.77 | 10.52 | 12.345 | —— | —— |

The parameter $A_0$ in Eq. (31) was determined by fitting the proposed failure criterion to the experimental data presented in Table 4. The dataset, which was first normalized, includes uniaxial compressive ($f_c$) and tensile ($f_t$) strengths, alongside ultimate biaxial stresses ($\sigma_1$ and $\sigma_2$). The fitting was performed by iteratively adjusting $A_0$ to minimize the deviation from the experimental points, while $\rho_s$ was held constant at its average value. This procedure yielded an optimal $A_0$ value of 1.57, resulting in a low root mean squared error (RMSE) of 0.0166. Fig. 6 visually confirms the good agreement between the fitted curve and the experimental data. Substituting the determined parameters into Eq. (31) gives the final expression for the equivalent stress:

$$\overline{\sigma}_{\mathrm{eq}} = 0.459 \cdot \overline{I}_1 + 0.062 \cdot \sqrt{132.87 \cdot \overline{I}_1^2 + 296.28 \cdot \overline{J}_2} \quad (52)$$

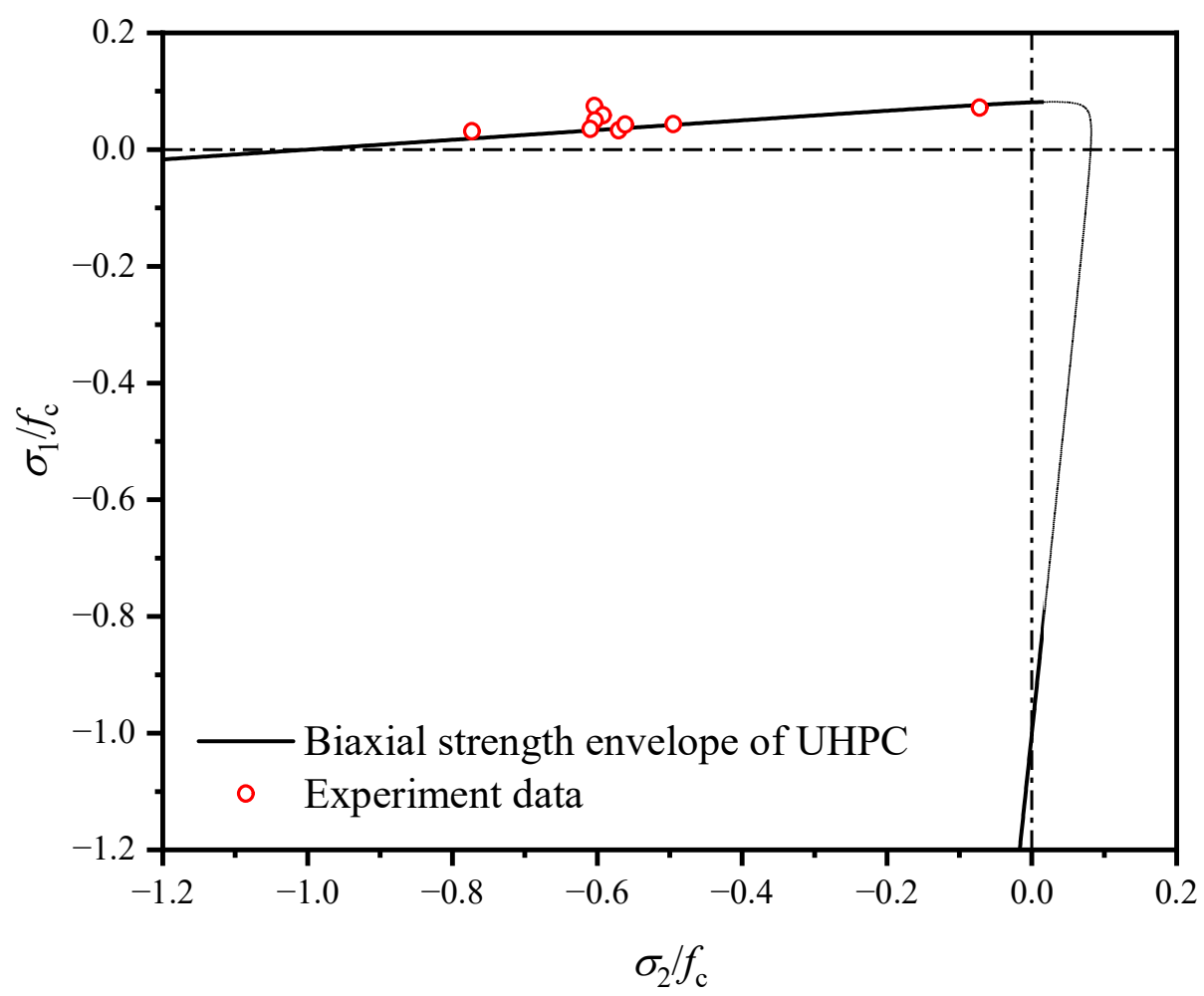

Fig. 6. Biaxial strength envelope of UHPC

## 3.2 Mode-I fracture and fatigue

### 3.2.1 Mode-I fracture

To evaluate the mode-I fracture behavior of UHPC, we validate the PF-CZM model against the three-point bending test on notched UHPC beams performed by de Almeida Cerqueira [28]. The beam dimensions are: height 100mm, width 100mm, and total length 400 mm, with a support span 350mm. The preset central notch of 15mm in height and 0.3mm in width is located at the bottom of the mid-span.

The overall arrangement of the UHPC notched beam is shown in Fig. 7. The beam was loaded by MTS hydraulic actuator with a clip gauge recoding the Crack Mouth Opening Displacement (CMOD). Material parameters such as Young's modulus $E_0$, Poisson's ratio $\nu$, tensile strength $f_t$, and fracture energy $G_f$ are calibrated based on experimental results during numerical simulation.

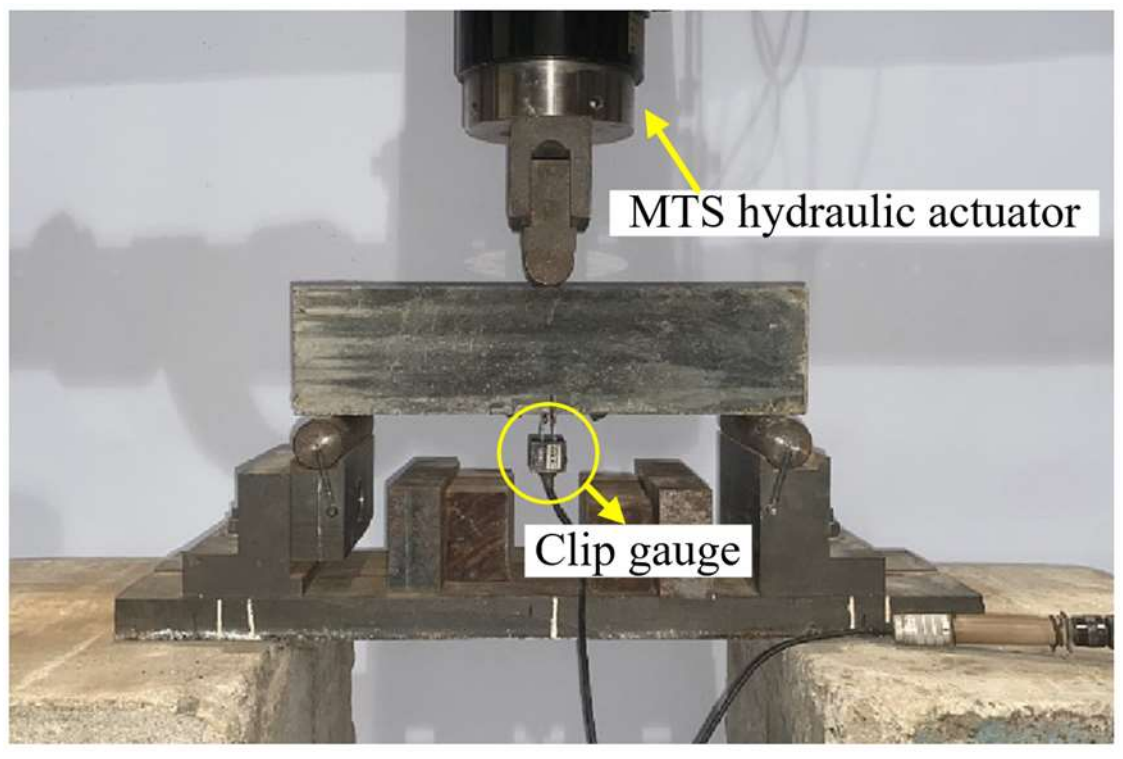

(a) The specimen and loading mode [28]

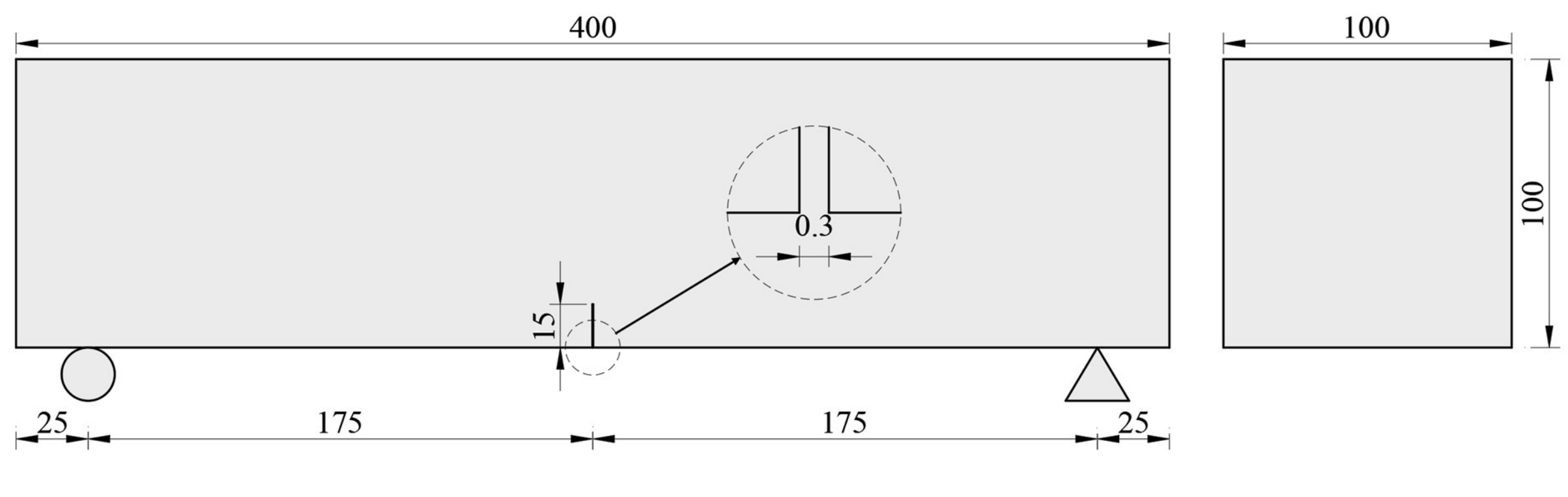

(b) Geometric dimensions

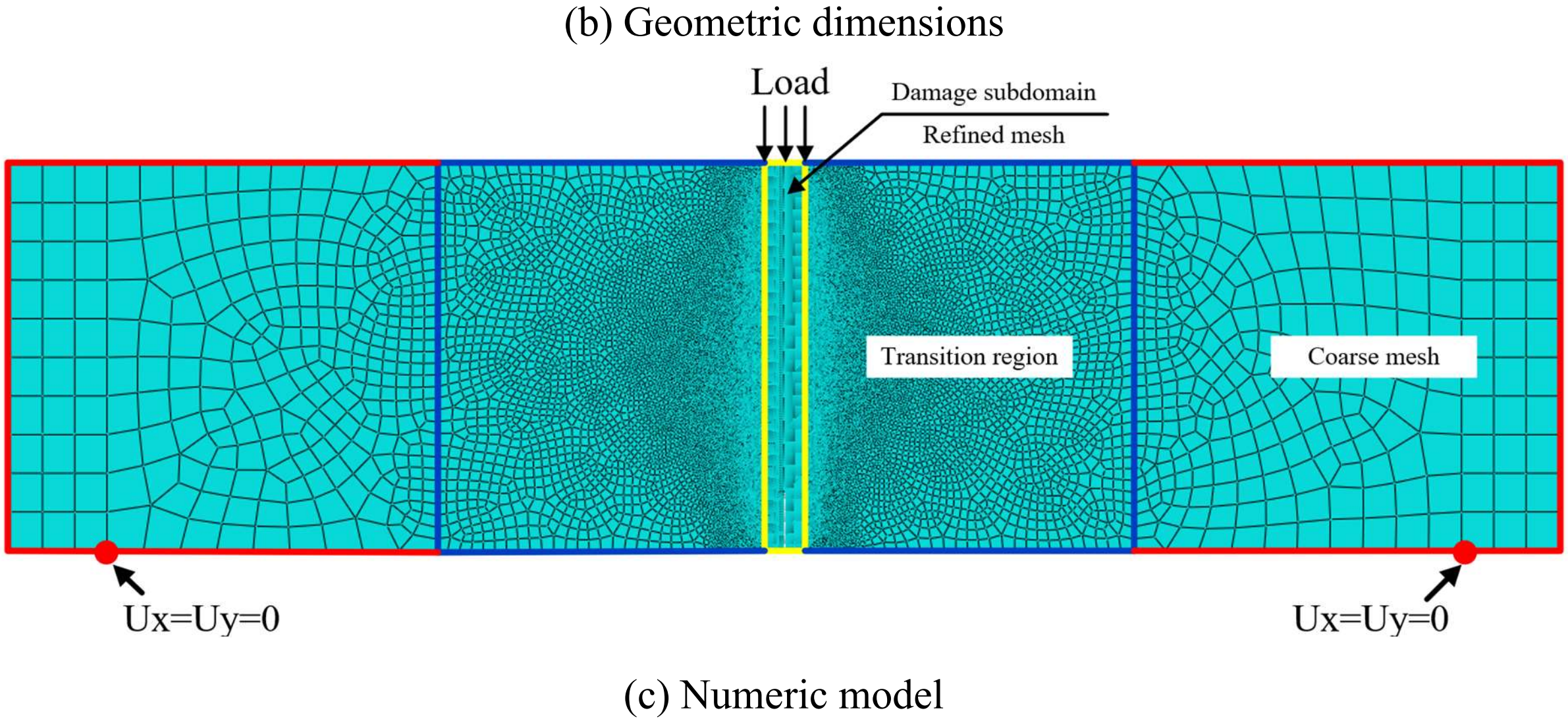

(c) Numeric model

Fig. 7. Overall arrangement and numeric model of UHPC notched beam

The boundary conditions and mesh of the numerical model are shown in Fig. 7 (c). The UHPC notched beam is modeled with the linear quadrilateral element CPE4T for temperature displacement coupling analysis. The two ends of the beam are simply supported, with the left node constrained the degrees of freedom in the $x$ and $y$ directions, and the right node constrained the degrees of freedom in the $y$ direction. Displacement load is applied at the mid span of the beam through a line load with a width of 10mm.

Unlike traditional PFM that heavily rely on the length scale $b$, in PF-CZM, the length scale $b$ is a numerical parameter rather than a material parameter. The smaller its value, the higher the accuracy of the crack area. When its value is less than a certain constant, it has almost no effect on the analysis results. Previous studies have shown that maintaining of the element size smaller than one-fifth of the length scale $b$ can ensure the mesh objectivity [9,38]. The refined mesh size will increase numerical computation costs. In order to balance computational accuracy and cost, a transition from refined to large mesh size is adopted. As shown in Fig. 7 (c), a small mesh size is used in the damage subdomain within 10mm of the mid span, which is the area where cracks initiate and propagate. A large mesh size is used within 180-400mm of the mid span, and the transition region of the mesh is within 10-180mm of the mid span. In order to investigate the influence of length scale $b$ on the analysis accuracy, four length scales $b$ are selected, namely 6mm, 4.5mm, 3mm, and 1.5mm, with corresponding mesh sizes of 1.2mm, 0.9mm, 0.6mm, and 0.3mm.

The material parameters are calibrated based on experimental results, including Young's modulus $E_0$ = 40000MPa, Poisson's ratio $\nu$ = 0.18, tensile strength $f_t$ = 6MPa, and fracture energy $G_f$ = 12N/mm. Due to the wide range of strength grades of UHPC and differences in fiber shape, fiber volume fraction, and other factors, significant variations occur in tensile strength and fracture energy across sections. As shown in Fig. 8 (a), the load-CMOD curves of numerical simulation with four different length scales $b$ are mostly consistent with the experimental results, with only a slight deviation in the load rise stage. As the length scale $b$ decreases, the accuracy of PF-CZM result continues to improve, converging toward experimental results. These findings confirm that the PF-CZM effectively simulate the fracture behavior of UHPC and the analysis results is insensitive to the length scale $b$, highlighting the superiority of PF-CZM over traditional PFM.

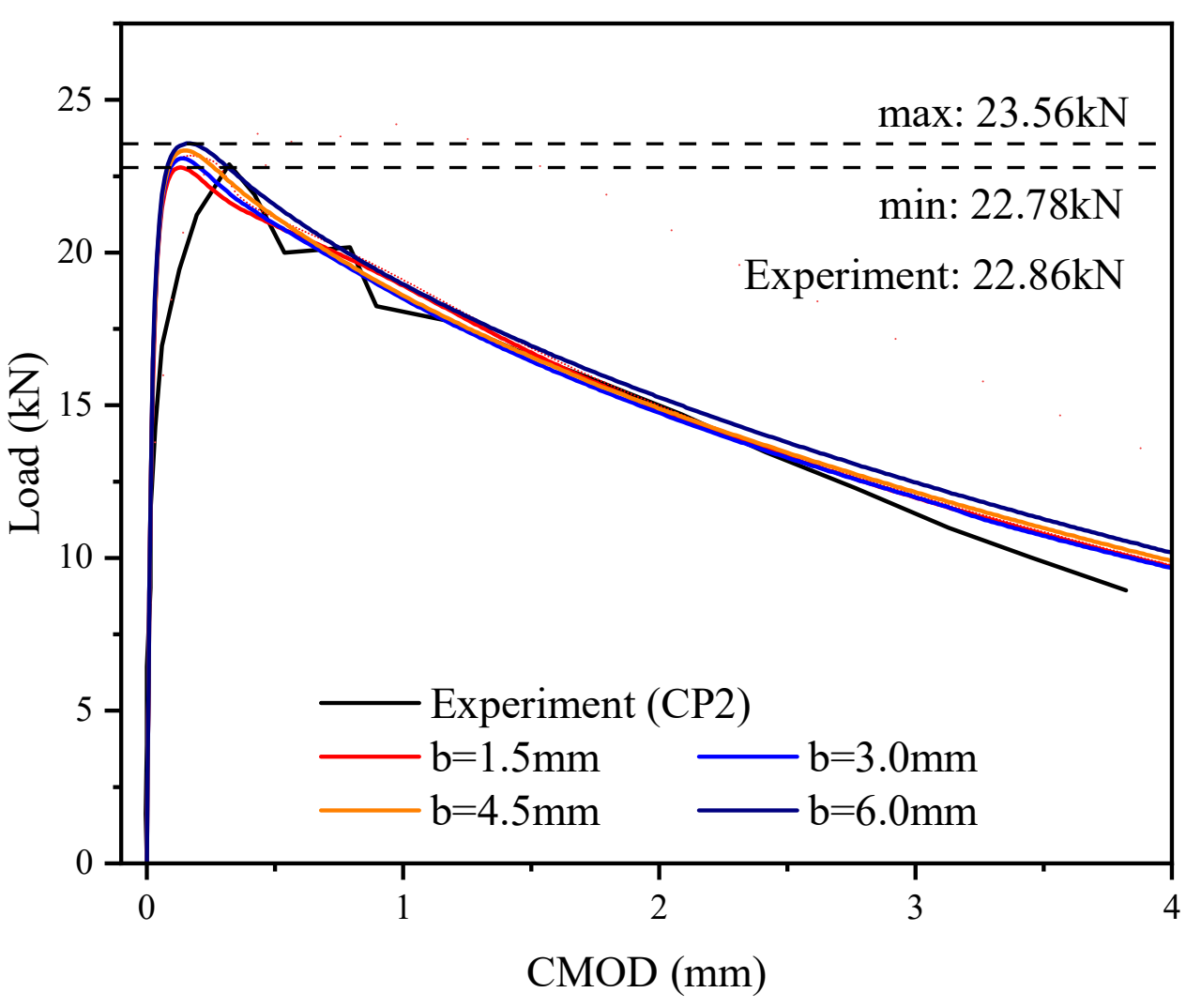

(a) Load-CMOD curves

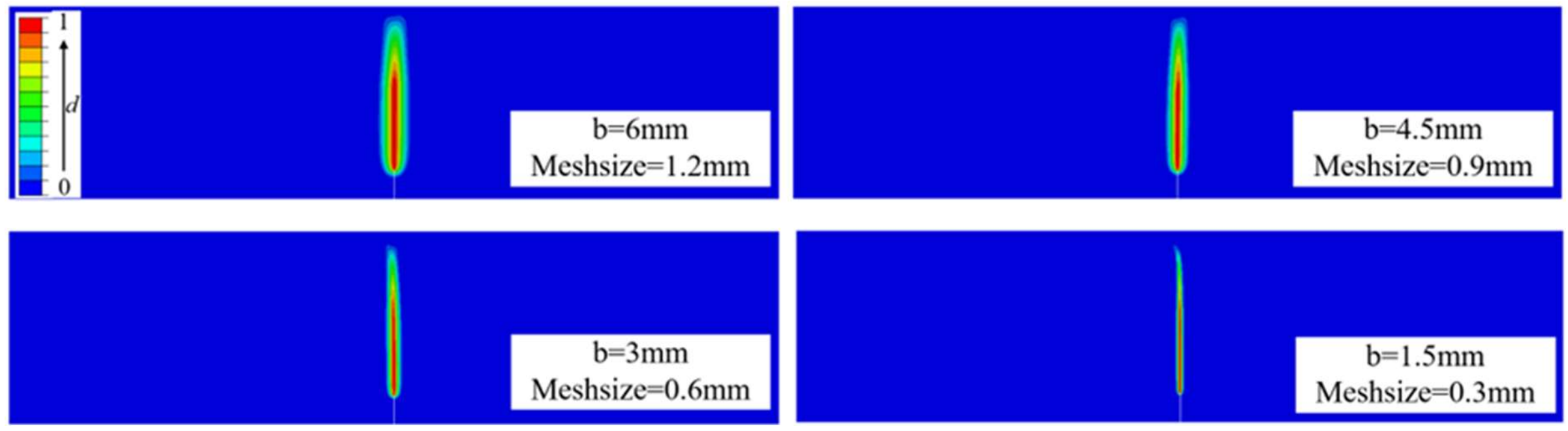

(b) Mode-I crack propagation

Fig. 8. Load-CMOD curves and mode-I crack propagation of the UHPC notched beam with different length scale *b*

Figure 8 (b) illustrates the simulated failure modes, highlighting the effect of the length scale parameter *b*. Cracks originate from the notches and propagate vertically upwards until the notched beams completely fracture. As the length scale *b* decreases, the width of the crack narrows, but the crack path is not affected. The results indicate that as length scale *b* decreases, the regularized crack converges towards an ideal sharp crack, and once again proves that the length scale *b* in PF-CZM affects the width of crack band rather than the crack mode.

3.2.2 Mode-I fatigue

To further validate the accelerated PF-CZM for fatigue, continuously, the three-point bending fatigue test of UHPC notched beams conducted by de Almeida Cerqueira [28] is selected. The fatigue test includes two load amplitudes, with upper limit loads of $S_{max}$ = $0.9P_{0.5mm}$ and $S_{max}$ = $0.8P_{0.5mm}$, and the load ratio is 0.3. $P_{0.5mm}$ is the bearing capacity of the UHPC notched beam when the CMOD reaches 0.5mm. Since these fatigue specimens originate from the same batch as those used in the monotonic tests in Section 3.2.1, their geometric dimensions and material properties are identical. Consequently, the material parameters used in accelerated PF-CZM for fatigue, such as tensile strength, fracture energy, etc., are the same as those in section 3.2.1.

As described in accelerated PF-CZM for fatigue in Section 2.2, the fatigue damage accumulation parameter $k_f$ governs the fatigue life of the material by modulating the fatigue damage accumulation rate. Therefore, four $k_f$ are examined for analysis, namely 5, 3, 2, 1, and 0.5. When the upper limit load $S_{max}$ is $0.9P_{0.5mm}$, the fatigue life range of the specimen is 252-2028 cycles, which belongs to low cycle fatigue, and cycle by cycle simulation is used for analysis. When the upper limit load $S_{max}$ is $0.8P_{0.5mm}$, the fatigue life of the specimen is greater than 700000 cycles, which belongs to high cycle fatigue, and the acceleration algorithm in Section 2.3 is used for analysis.

As shown in Fig. 9, when the upper limit load $S_{max}$ is $0.9P_{0.5mm}$, as the fatigue damage accumulation parameter $k_f$ decreases, the fatigue life of the specimen continues to increase. When $k_f = 1$, the fatigue life obtained by the PF-CZM for fatigue closely matches to the test.

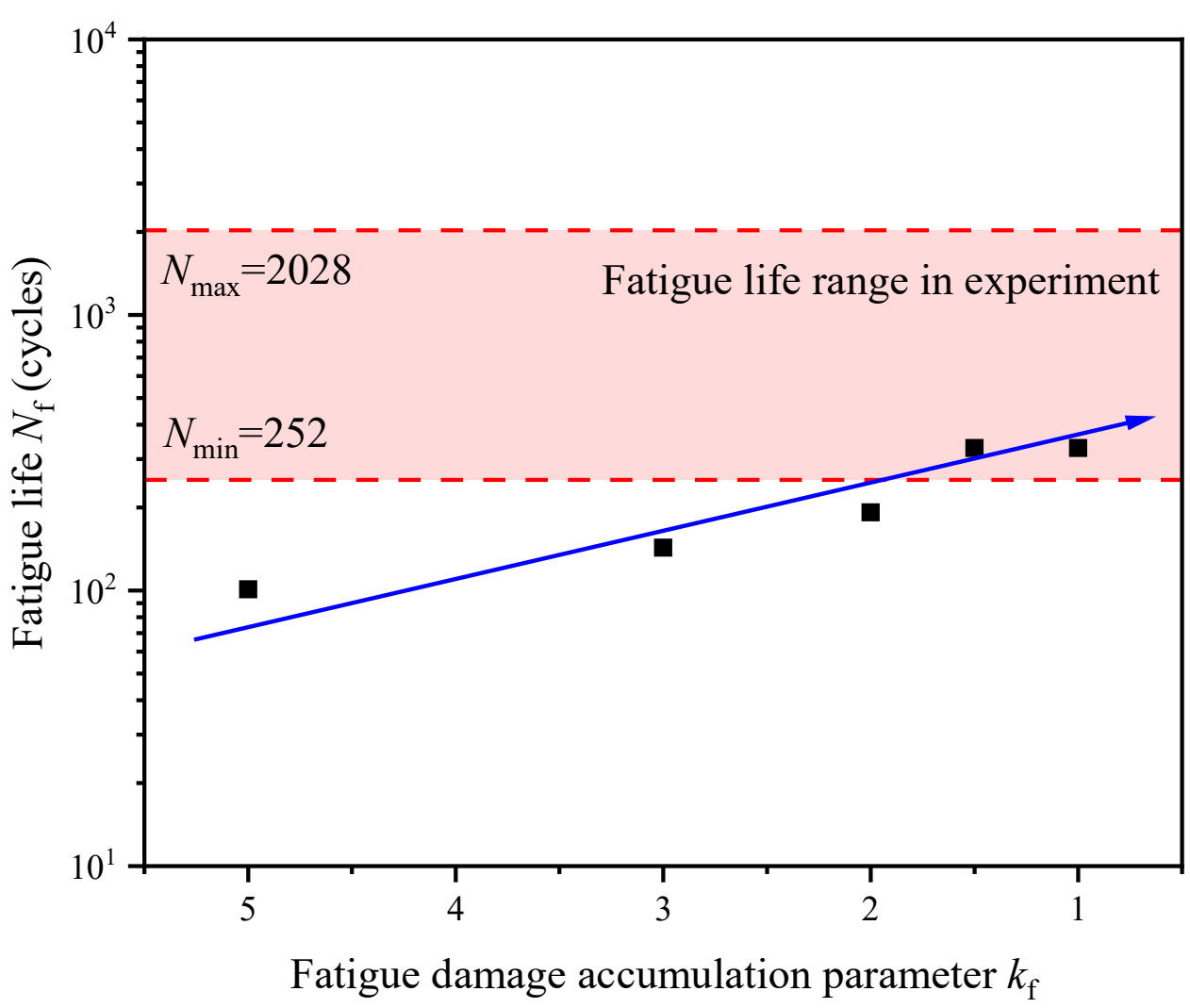

Fig. 9. Fatigue life of UHPC notched beam with different $k_f$ ($S_{max} = 0.9P_{0.5mm}$)

When the upper limit load $S_{max}$ is $0.9P_{0.5mm}$, as $k_f$ is reduced constantly, the CMOD of the specimen also decreases, but both are close to the test as presented in Fig. 10 (a). Fig. 10 (b) illustrates similar relations as Fig.10 (a). When the upper limit load $S_{max}$ is $0.8P_{0.5mm}$, the CMOD of the specimen decreases as $k_f$ is reduced constantly. When $k_f = 1$, the CMOD of the specimen closely match the test.

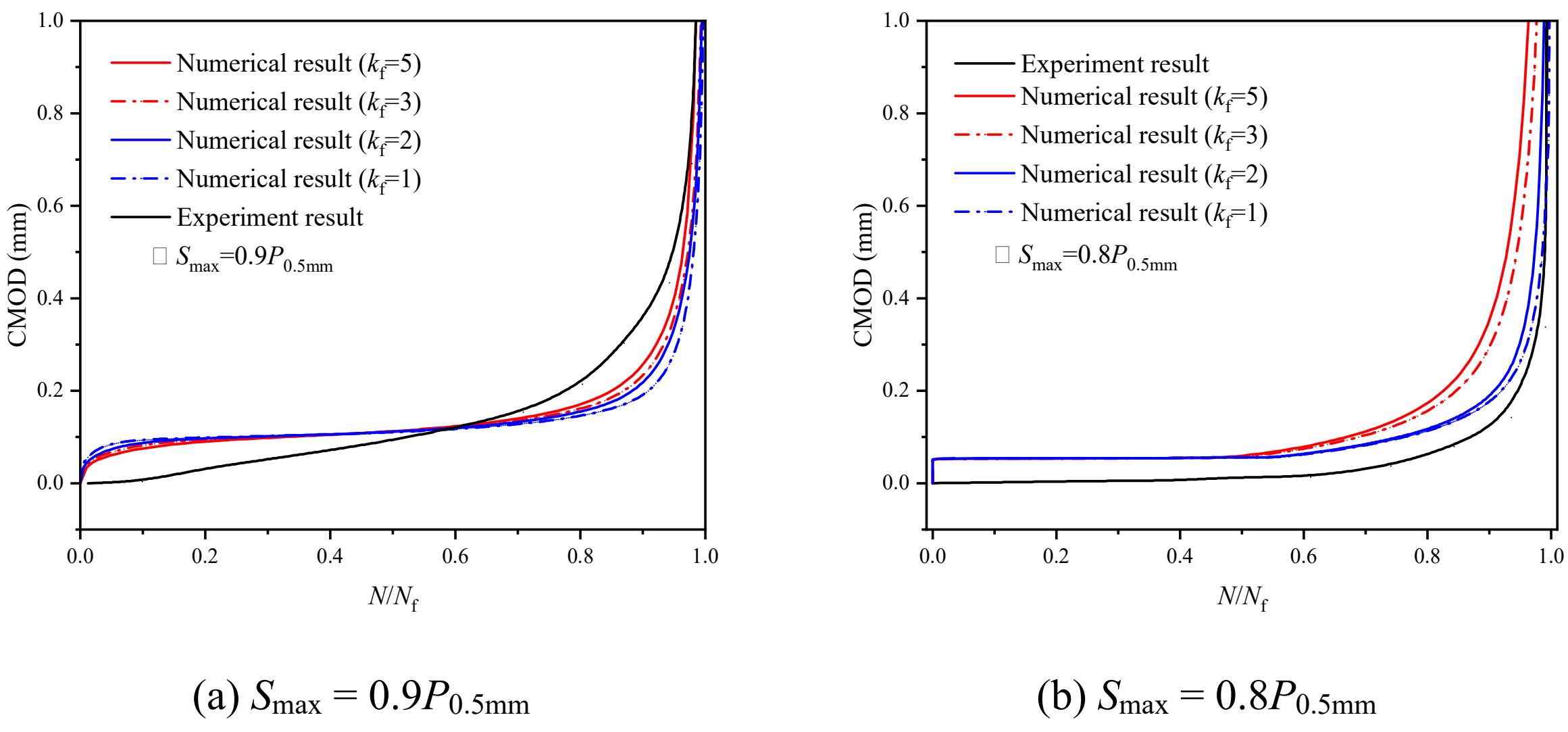

(a) $S_{max} = 0.9P_{0.5mm}$ (b) $S_{max} = 0.8P_{0.5mm}$

Fig. 10. Fatigue creep curves of UHPC notched beam with different $k_f$

The fatigue crack path of UHPC notched beams under cycle-by-cycle simulation and acceleration algorithm in PF-CZM is shown in Fig. 11. It should be specially noted that only one figure ($S_{max}$ = 0.8$P_{0.5mm}$) is presented in Fig.11, since the two approaches yield identical crack propagation. Both the identical results and the crack propagation reflect the accuracy of accelerated PF-CZM for UHPC fatigue analysis. Moreover, because the influence of fiber volume fraction on the fatigue strength of UHPC, a specific $k_f$ can't reflect the fatigue behavior of UHPC across different fiber volume fractions. However, the value of $k_f$ given in this section could be referred and adjusted appropriately through tests for UHPC with different fiber volume fractions.

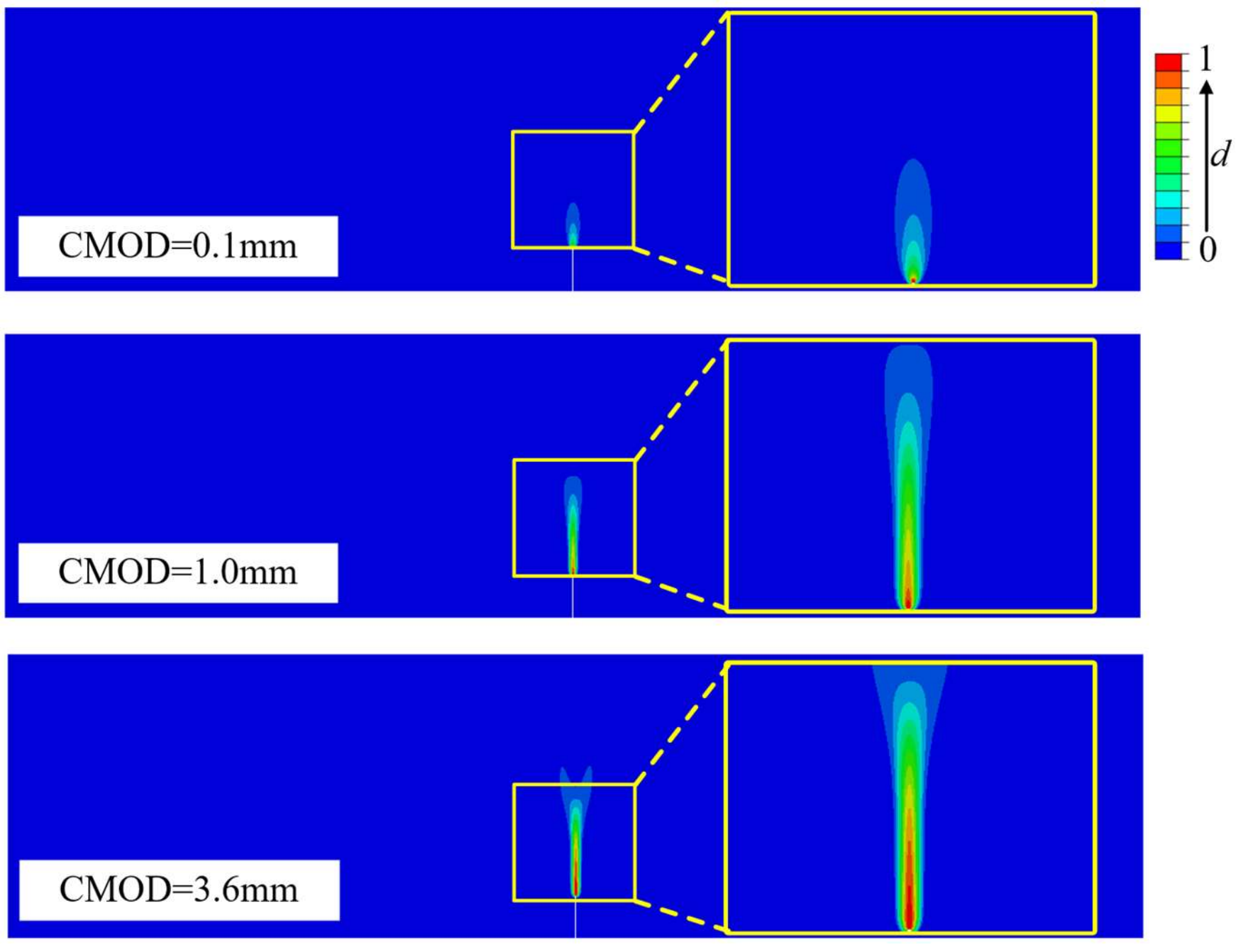

Fig. 11. Mode-I fatigue crack propagation ($S_{max}$ = 0.8$P_{0.5mm}$)

## 3.3 Mixed-mode fracture and fatigue

### 3.3.1 Mixed-mode fracture

To demonstrate the proposed model's capability in capturing mixed-mode fracture behavior, we analyze a UHPC notched beam featuring a 100 mm offset from the midspan. The beam dimensions are: height 100mm, width 100mm, length 500mm and the calculated span 400mm. The preset notch geometry is: height 50mm and width 5mm. The overall arrangement of the UHPC notched beam is shown in Fig. 12 (a). The material parameters such as Young's modulus $E_0$, Poisson's ratio $\nu$, tensile strength $f_t$, and fracture energy $G_f$ are identical as the calibration values in section 3.2. Mixed-mode fracture and fatigue simulations are performed as predictive analyses based on the calibrated material parameters in section 3.2.

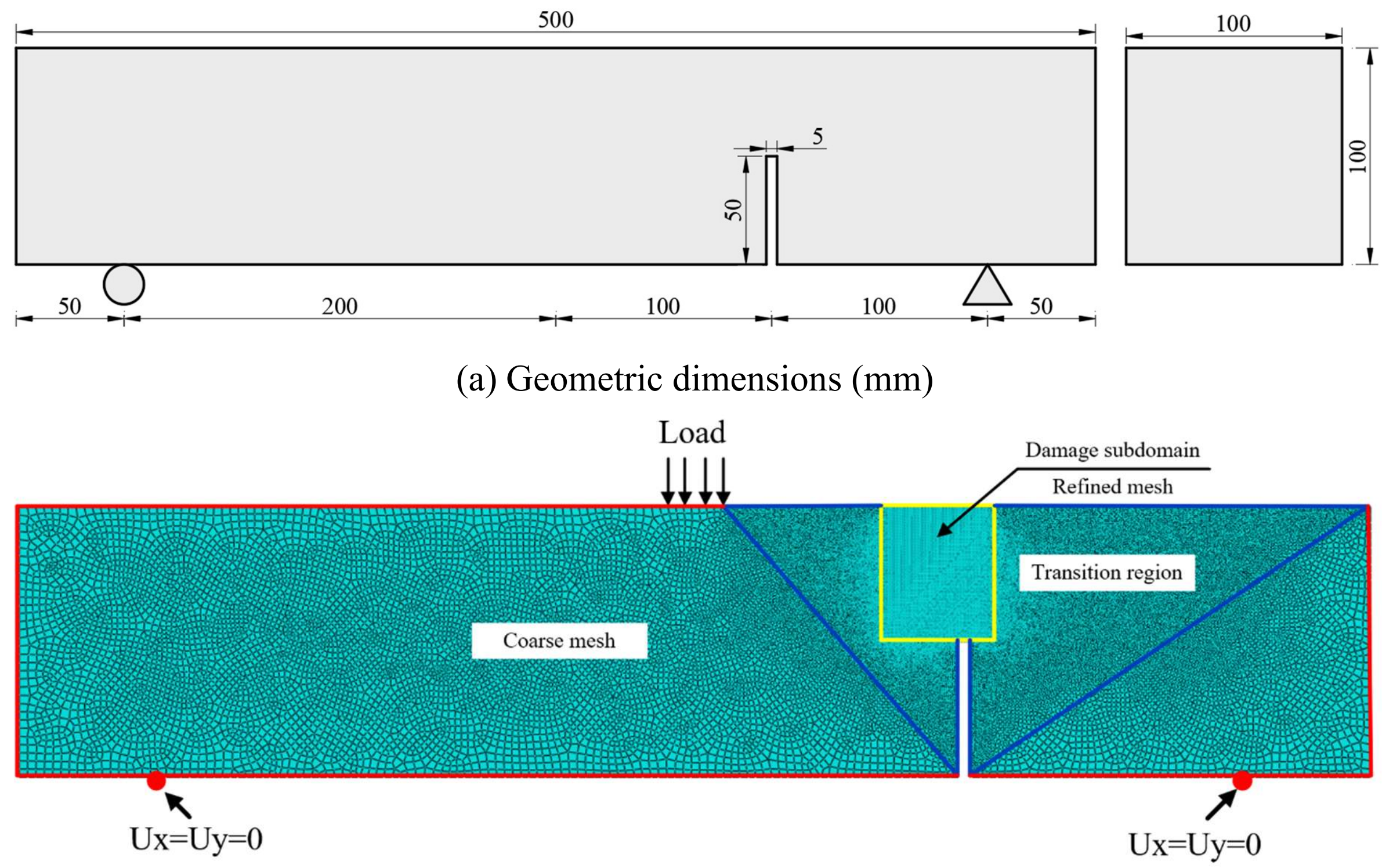

(a) Geometric dimensions (mm)

(b) Boundary conditions and mesh division

Fig. 12. Overall arrangement of the UHPC notched beam

The boundary conditions and mesh division of the numerical model are shown in Fig. 12 (b). The UHPC notched beam is modeled using the linear quadrilateral element CPE4T for temperature displacement coupling analysis. The two ends of the beam are simply supported, with the left node constrained the degrees of freedom in the $x$ and $y$ directions, and the right node constrained the degrees of freedom in the $y$ direction. Displacement load is applied at the mid span of the beam through a line load with a width of 20mm. Similar to section 3.2, in order to balance computational accuracy and cost, a transition from refined to large mesh size is adopted as shown in Fig. 12 (b). Four length scales $b$ are examined for analysis, namely 6mm, 4.5mm, 3mm, and 1.5mm, with corresponding mesh sizes of 1.2mm, 0.9mm, 0.6mm, and 0.3mm.

As shown in Fig. 13 (a), the load-CMOD curves of the UHPC notched beam with four different length scales $b$ are basically consistent, indicating that PF-CZM is insensitive to the length scale $b$. The mixed-mode fracture of the UHPC notched beam with four different length scales $b$ are basically the same as compared in Fig. 13 (b). The crack originates from the notch and propagate to the upper left until the beam completely fracture. The results are similar as the Mode-I fracture, in which as length scale $b$ decreases the regularized crack converges towards an ideal sharp crack.

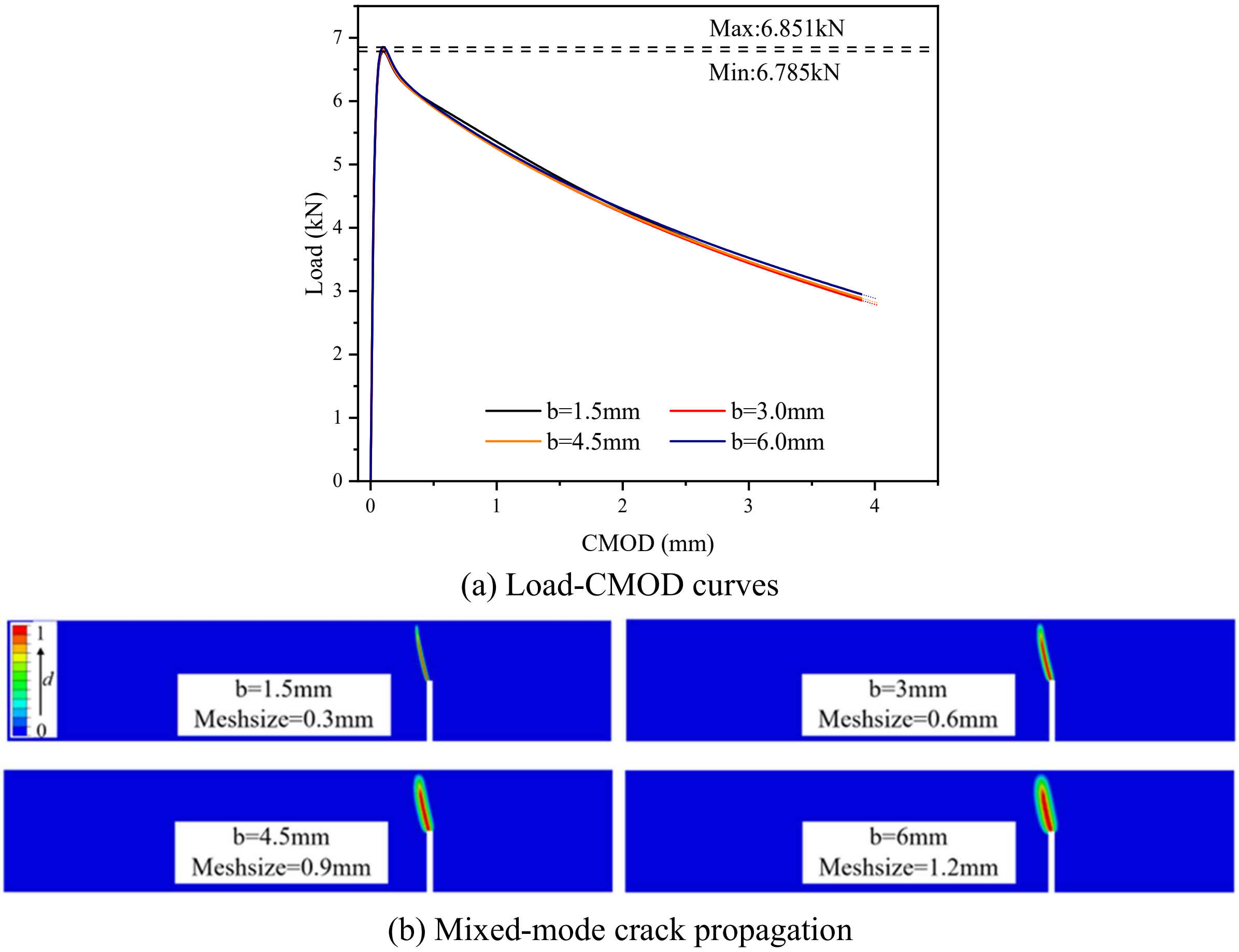

(a) Load-CMOD curves

(b) Mixed-mode crack propagation

Fig. 13. Load-CMOD curves and mixed-mode crack propagation of the UHPC notched beam with different *b*

3.3.2 Mixed-mode fatigue

To further showcase the model's capability in the mixed-mode fatigue behavior of UHPC, the same UHPC notched beam as in section 3.2.1 is employed for fatigue analysis, with identical geometric dimensions and material parameters, and the fatigue damage accumulation parameter $k_f$=1. Three load amplitudes are set as: the upper limit loads being $S_{max}$=0.9$P_{max}$, $S_{max}$ = 0.8$P_{max}$ and $S_{max}$ = 0.7$P_{max}$, and the lower limit load $S_{min}$ = 0. $P_{max}$ represents the ultimate bearing capacity of the UHPC notched beams. As shown in Fig. 14 (a), the CMOD of the UHPC notched beam increases with the increase of the load amplitude. The fatigue crack path of the UHPC notched beam ($S_{max}$ = 0.9$P_{max}$) is shown in Fig. 14 (b). When $S_{max}$ = 0.8$P_{max}$ and $S_{max}$ = 0.7$P_{max}$, the crack paths are similar.

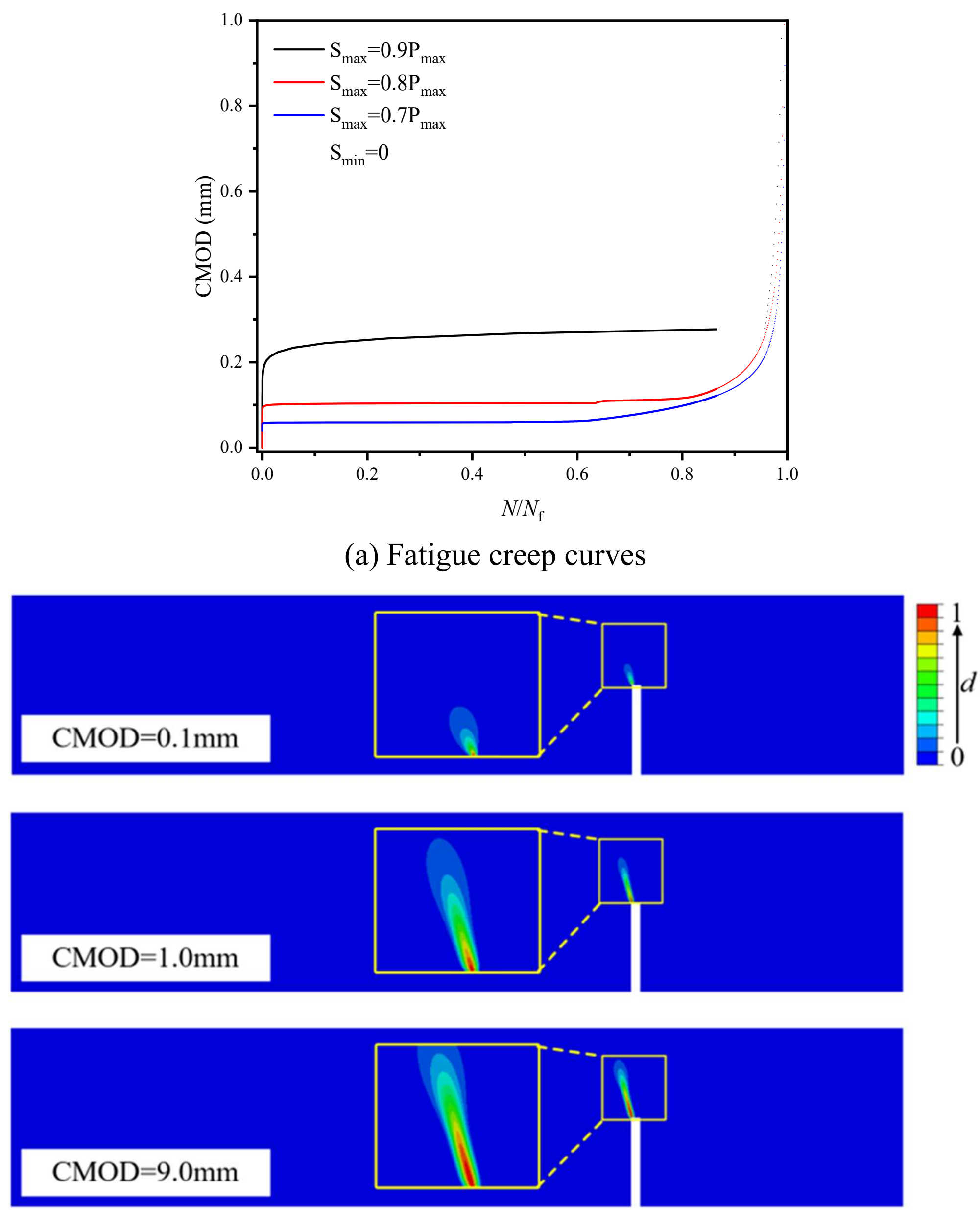

(a) Fatigue creep curves

(b) Mixed-mode fatigue crack propagation

Fig. 14. Fatigue creep curves with different load amplitudes and mixed-mode fatigue crack propagation ($S_{\max} = 0.9P_{\max}$) of UHPC notched beam

# 4 HSS fracture modeling

## 4.1 PF-CZM for HSS

This section adapts the proposed model to simulate the low-temperature brittle fracture of High-Strength Steel (HSS). The failure mode of steel structure is a critical concern in cold environments, where the material's fracture energy decreases significantly. Adapting the model requires specifying two key constitutive functions: the crack geometry function, and the energy degradation function. Firstly, the crack geometry function $\alpha(d) = 2d - d^2$ has been validated as appropriate for this context [9]. For the energy degradation function, previous study has shown that linear softening effectively characterizes brittle fracture behavior [51], as written in Eq. (53). The energy degradation function is provided in Eq. (54).

$$\sigma(w) = f_{\mathrm{t}} \max[1 - f_{\mathrm{t}} / (2G_{\mathrm{f}})w, 0] \quad (53)$$

$$\omega(d) = \frac{(1-d)^2}{(1-d)^2 + a_1 d \cdot (1 - d/2)} \quad (54)$$

For mode-I fracture of HSS, the crack driving force could be obtained by the first principal stress. While for mixed-mode fracture, the crack driving force could be derived by Eq. (30) following the generic stress-based failure criterion. When $A_0 = 0$ and $\rho_{\mathrm{s}} = 1$, the generic stress-based failure criterion is restored to the von Mises failure criterion [41], as expressed in Eq. (55), which is used to calculate the crack driving force for mixed-mode fracture of HSS.

$$\overline{\sigma}_{\mathrm{eq}} = \sqrt{3\overline{J}_2} \quad (55)$$

## 4.2 Mode-I fracture

To evaluate the Mode-I fracture behavior of HSS, we validate the PF-CZM model against the Center Notch Tension (CNT) test of HSS conducted by Meissne [33]. The CNT specimen's dimension is: height 400mm, width 40mm, length 2500mm; and the preset crack length at the center is 60mm as shown in Fig. 15 (a). The CNT was loaded by a hydraulic actuator, and the displacement was measured through displacement transducers. The tensile strength of HSS is $f_{\mathrm{t}}$ =1045MPa, and other material parameters such as Young's modulus $E_0$, Poisson's ratio $\nu$, and fracture energy $G_{\mathrm{f}}$ are calibrated through test results during numerical simulation.

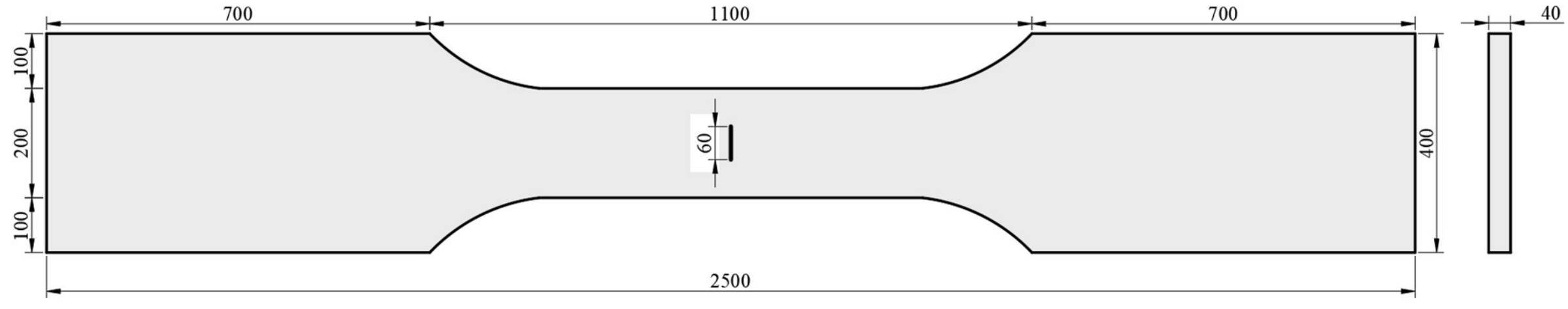

(a) Geometric dimensions

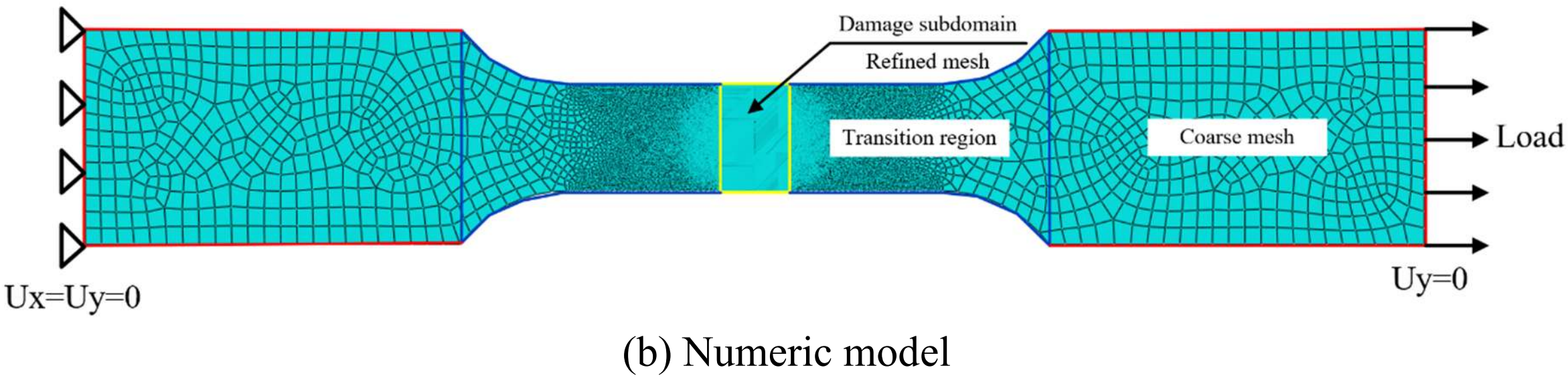

(b) Numeric model

Fig. 15. Geometric dimensions and numeric model of the CNT

The boundary conditions and mesh division of the numerical model are shown in Fig. 15 (b). The CNT is modeled using the linear quadrilateral element CPE4T for temperature displacement coupling analysis. Uniaxial tensile loading is adopted, the degrees of freedom in the $x$ and $y$ directions are constrained on the left, the degree of freedom in the $y$ direction is constrained on the right, and displacement loading is applied in the $x$ direction on the right. The transition from small to large mesh size is adopted, with a mesh size of 0.4mm at the damaged subdomain.

The material parameters are calibrated based on test results, including the Young's modulus of $E_0 = 237900$MPa, Poisson's ratio of $\nu = 0.2$, and fracture energy of $G_f = 122$N/mm. The numerical simulation results agree well with the test data as presented in Fig. 16 (a). The crack propagation path of CNT is shown in Fig. 16 (b), and it is observed that the numerical simulation results are consistent with the test phenomena. These findings indicate that the PF-CZM model effective captures the low-temperature brittle fracture behavior of HSS.

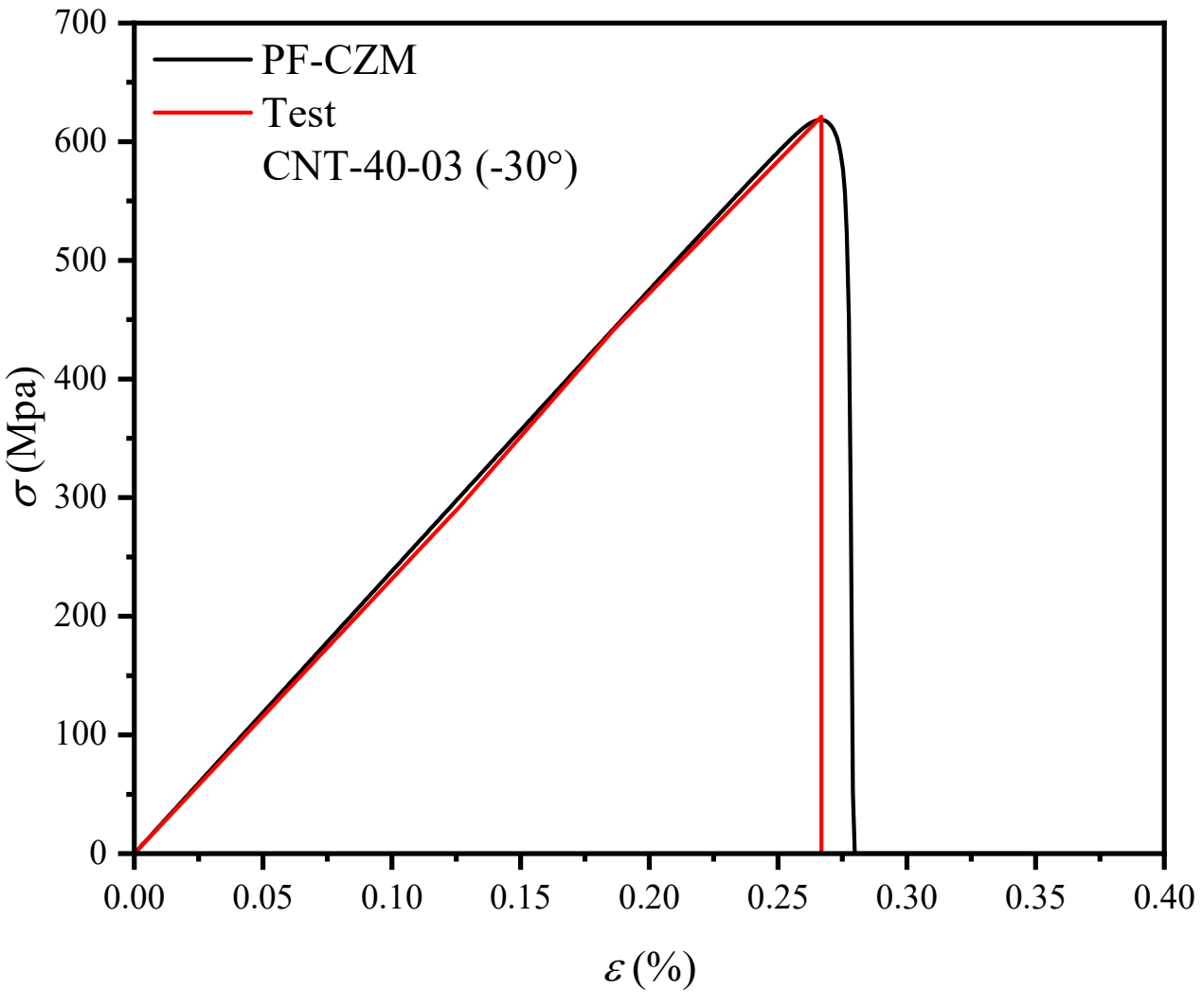

(a) Stress-strain curves

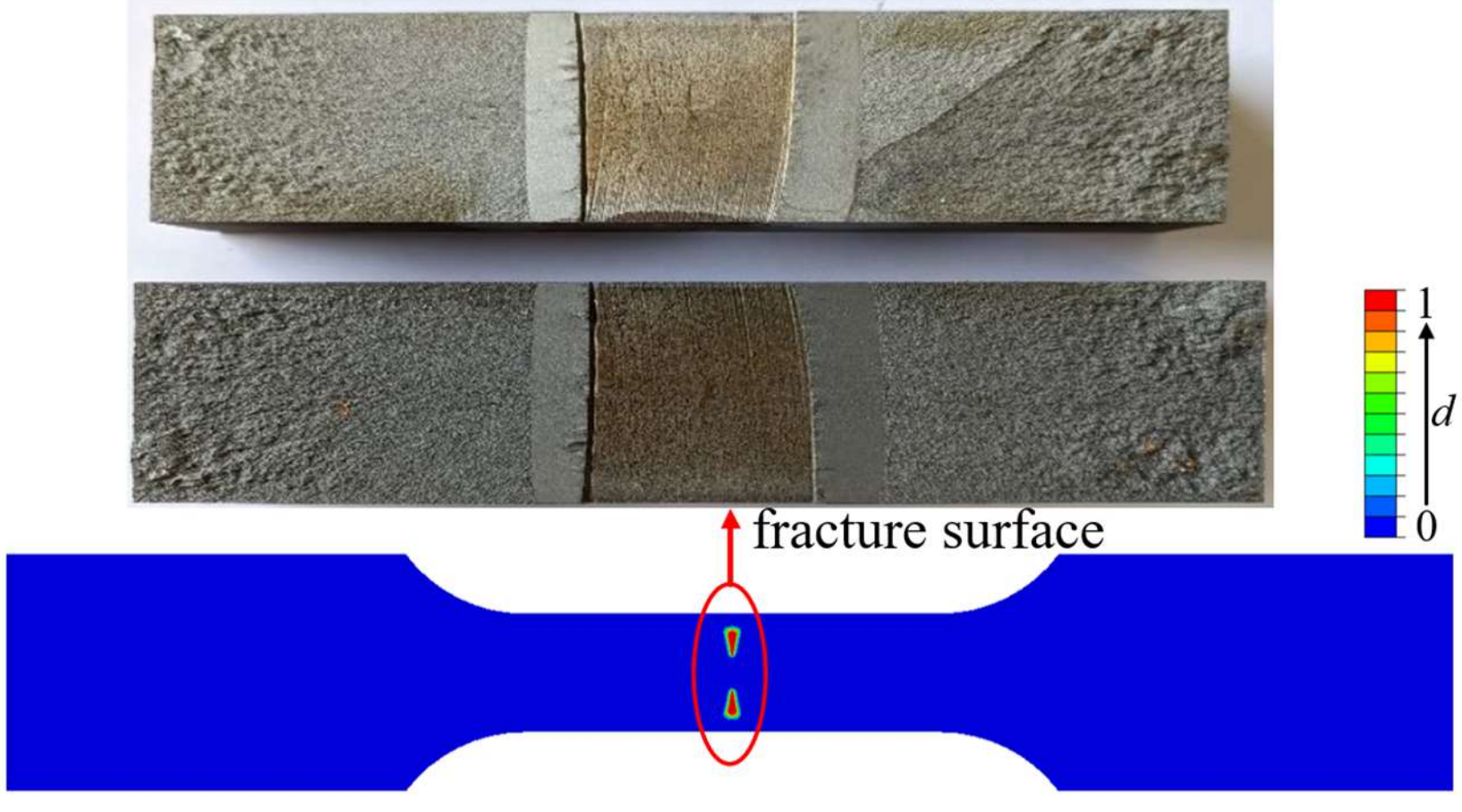

(b) Crack propagation path

Fig. 16. Stress-strain curves and crack propagation path of the CNT [33]

## 4.3 Mixed-mode fracture

We furtherly investigate the mixed-mode fracture behavior of HSS, and Compact Tension Specimen (CTS) is selected for analysis. The dimension of CTS is: height 120mm, width 90mm, and thickness 10mm. The preset crack length is 45mm and the width is 1mm. The detail geometry of CTS is shown in Fig. 17 (a). The CTS is subjected to tension-shear coupling load using the loading device shown in Fig. 17 (b) The material parameters such as Young's modulus $E_0$, Poisson's ratio $\nu$, tensile strength $f_t$, and fracture energy $G_f$ are the same as those in section 4.2. Mixed-mode fracture simulation is performed as predictive analyses based on the calibrated material parameters in section 4.2.

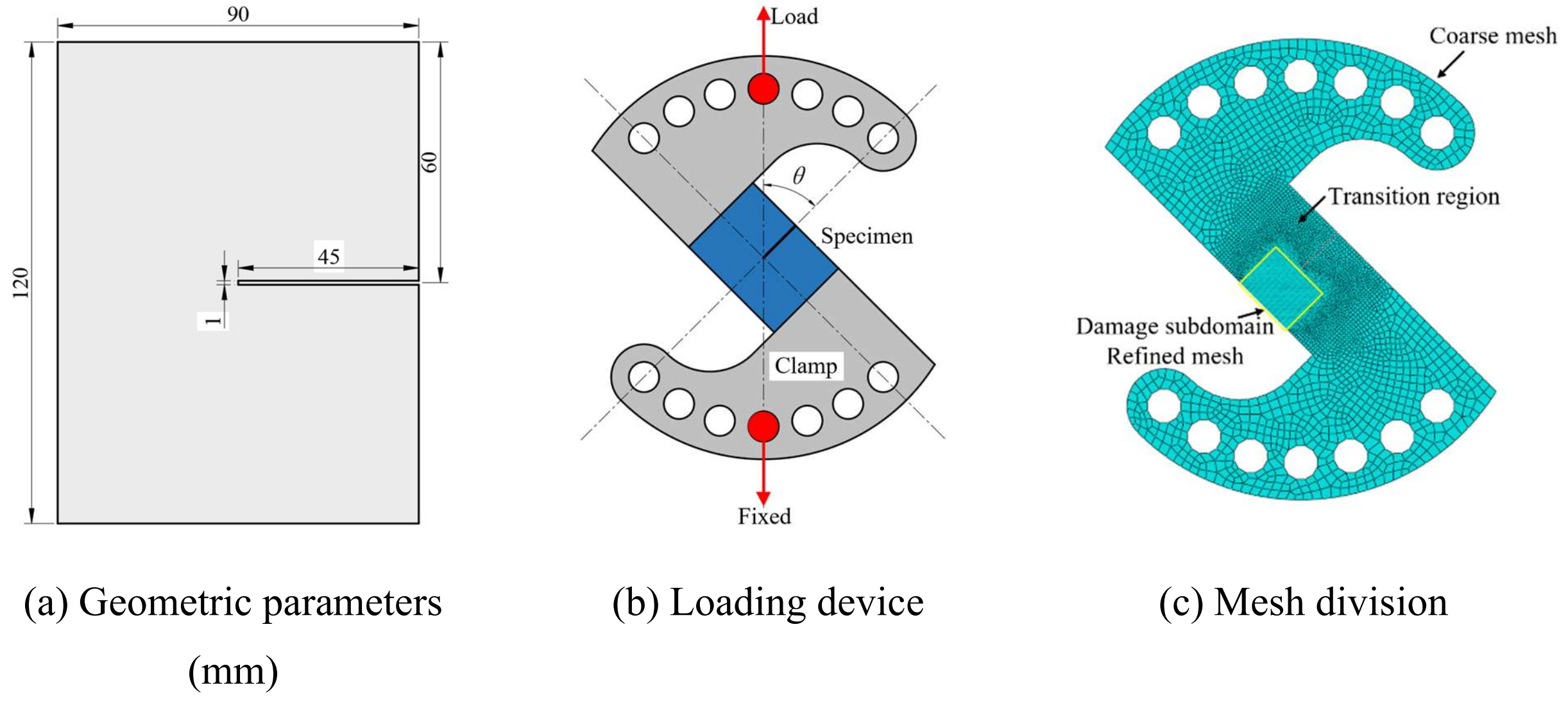

(a) Geometric parameters (mm)　　(b) Loading device　　(c) Mesh division

Fig. 17. Geometry and numeric model of the CTS

The mesh division of the numerical model are shown in Fig. 17 (c). The CTS is modeled using the linear quadrilateral element CPE4T for temperature displacement coupling analysis. The lower end of the CTS is fixed, while the upper end is loaded with displacement at loading angles $\theta$ of 15 °, 30 °, 45 °, 60 °, and 75 °, respectively. Transition from small to large mesh size is adopted, with a mesh size of 0.4mm in the damage subdomain. As shown in Fig. 18 (a), the load displacement curves of the CTS under different loading angles present that the maximum load decreases with the decrease of the loading angle.

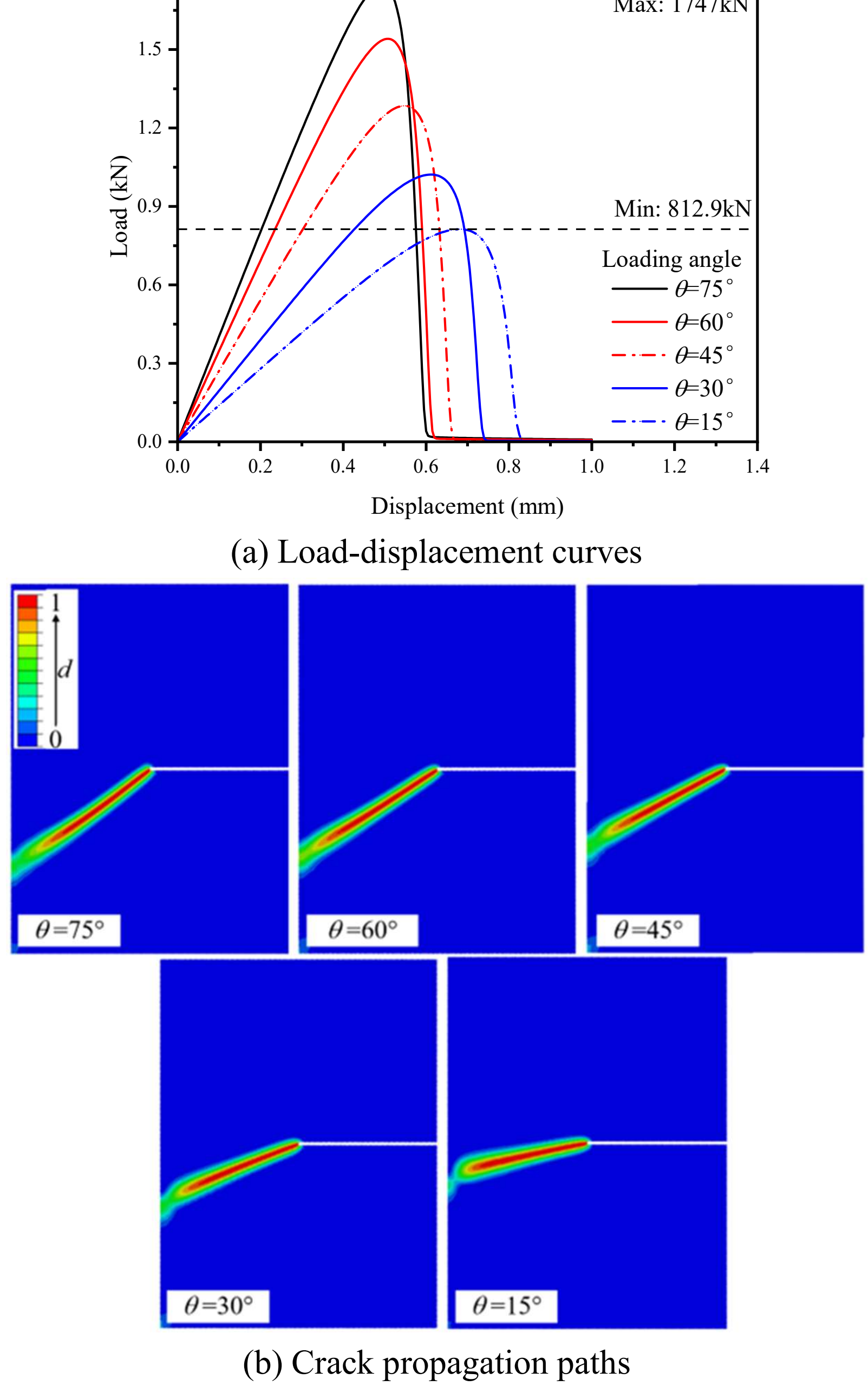

(a) Load-displacement curves

(b) Crack propagation paths

Fig. 18. Load-displacement curves and crack propagation paths of the CTS with different loading angles

The crack propagation path of the CTS gradually approaches horizontal as the loading angle decreases as demonstrated in Fig. 18 (b). As the loading angle decreases, shear stress on the crack surface increases while normal stress decreases. Consequently, crack propagation becomes shear-dominated, progressing toward a horizontal (pure-shear) orientation. The PF-CZM predictions accurately capture this transition.

## 5. Conclusions

In this study, a Phase-Field Cohesive Zone Model (PF-CZM) is employed to model the fracture and high-cycle fatigue behavior of Ultra-High Performance Concrete (UHPC) and High Strength Steel (HSS), incorporating a fatigue degradation function and an acceleration algorithm. The main conclusions are as follows:

- The energy degradation function applicable to UHPC is analyzed based on the uniaxial tensile test results of UHPC. The constitutive model of UHPC in the PF-CZM framework is established, achieving effective simulation of I-mode fracture of UHPC.
- A generic stress-based failure criterion for UHPC, *i.e.*, the hyperbolic biaxial strength envelope, is derived. With this criterion, it opens up the precise calculation of the crack driving force and effective simulation of mixed-mode fracture of UHPC.
- A fatigue degradation function and an acceleration algorithm are integrated into PF-CZM, and the fatigue damage accumulation parameter is calibrated through test results, enabling effective simulation of mode-I and mixed-mode high cycle fatigue of UHPC.
- The PF-CZM framework is successfully extended to HSS. Precise reproducing of mode-I and mixed-mode fracture of HSS at low temperatures is achieved using PF-CZM.

## Disclosure statement

No potential conflict of interest was reported by the author(s).

## Funding Declaration

This work was supported by National Natural Science Foundation of China (Grant No.

51908422).

## References

[1] de Jesus, A. M., Matos, R., Fontoura, B. F., Rebelo, C., da Silva, L. S., & Veljkovic, M. (2012). A comparison of the fatigue behavior between S355 and S690 steel grades. *Journal of constructional steel research*, *79*, 140-150.

[2] Xiong, X., Wu, M., Shen, W., Li, J., Zhao, D., Li, P., & Wu, J. (2022). Performance and microstructure of ultra-high-performance concrete (UHPC) with silica fume replaced by inert mineral powders. C*onstruction and Building Materials*, 327, 126996.

[3] Bencherif, N., Slamene, A., Hamza, B., Trari, I. I., Boudaib, I., & Mokhtari, M. (2025). Predictive plasticity unveiled: XFEM modeling of cyclic failure in pressurized straight pipelines under three-point bending. *Mechanics Based Design of Structures and Machines*, *53*(5), 3547-3571.

[4] Li, J., Xiong, Z., Fan, H., & Liu, H. (2025). Conceptual design and axial-flexural performance of composite dowels columns. *Journal of Building Engineering*, 114275.

[5] Azevedo, C. R. D. F., Magarotto, D., Araújo, J. A., & Ferreira, J. L. A. (2009). Bending fatigue of stainless steel shear pins belonging to a hydroelectric plant. *Engineering Failure Analysis*, 16(4), 1126-1140.

[6] Xiong, Z., Li, J., Wolters, K., Mou, X., & Feldmann, M. (2025). Fatigue performance of composite dowels under combination of shear and tension: Part I–Steel component. *Steel Construction*, 18(1), 52-62.

[7] Francfort, G. A., & Marigo, J. J. (1998). Revisiting brittle fracture as an energy minimization problem. *Journal of the Mechanics and Physics of Solids*, *46*(8), 1319-1342.

[8] Li, P., Li, W., Li, B., Yang, S., Shen, Y., Wang, Q., & Zhou, K. (2023). A review on phase field models for fracture and fatigue. Engineering Fracture Mechanics, 289, 109419.

[9] Wu, J. Y. (2017). A unified phase-field theory for the mechanics of damage and quasi-brittle failure. *Journal of the Mechanics and Physics of Solids*, *103*, 72-99.

[10] Wang, Q., Feng, Y. T., Zhou, W., Cheng, Y., & Ma, G. (2020). A phase-field model for mixed-mode fracture based on a unified tensile fracture criterion. *Computer Methods in Applied Mechanics and Engineering*, *370*, 113270.

[11] Shlyannikov, V., Tumanov, A., & Khamidullin, R. (2025). Distinctive Features of Crack Growth Under Cyclic Loading at Elevated Temperature in the Context of Phase-Field

Fracture. *Fatigue & Fracture of Engineering Materials & Structures*.

[12] Hou, J., Zhang, Y., Tan, J., Peng, X., Li, Q., & Wang, C. (2025). Investigation of the Thickness Effects on Three-Dimensional Fracture Toughness in Metallic Materials via the Phase-Field Model. *Fatigue & Fracture of Engineering Materials & Structures*, *48*(3), 1215-1235.

[13] Wang, Q., Zhou, W., & Feng, Y. T. (2020). The phase-field model with an auto-calibrated degradation function based on general softening laws for cohesive fracture. *Applied Mathematical Modelling*, *86*, 185-206.

[14] Alessi, R., Vidoli, S., & De Lorenzis, L. (2018). A phenomenological approach to fatigue with a variational phase-field model: The one-dimensional case. *Engineering fracture mechanics*, *190*, 53-73.

[15] Carrara, P., Ambati, M., Alessi, R., & De Lorenzis, L. (2020). A framework to model the fatigue behavior of brittle materials based on a variational phase-field approach. *Computer Methods in Applied Mechanics and Engineering*, *361*, 112731.

[16] Schreiber, C., Müller, R., & Kuhn, C. (2021). Phase field simulation of fatigue crack propagation under complex load situations. *Archive of Applied Mechanics*, *91*(2), 563-577.

[17] Schreiber, C., Kuhn, C., Müller, R., & Zohdi, T. (2020). A phase field modeling approach of cyclic fatigue crack growth. *International Journal of Fracture*, *225*, 89-100.

[18] Yan, S., Schreiber, C., & Müller, R. (2022). An efficient implementation of a phase field model for fatigue crack growth. *International Journal of Fracture*, *237*(1), 47-60.

[19] Yan, S., Müller, R., & Ravani, B. (2024). Thermomechanical fatigue life simulation using the phase field method. *Computational Materials Science*, *235*, 112829.

[20] Schreiber, C. (2021). *Phase field modeling of fracture: Fatigue and anisotropic fracture resistance*. Technische Universität Kaiserslautern.

[21] Ding, J., Yu, T., Fang, W., & Natarajan, S. (2024). An adaptive phase field modeling of fatigue crack growth using variable-node elements and explicit cycle jump scheme. Computer Methods in Applied Mechanics and Engineering, 429, 117200.

[22] Heinzmann, J., Carrara, P., Ambati, M., Mirzaei, A. M., & De Lorenzis, L. (2024). An adaptive acceleration scheme for phase-field fatigue computations. arXiv preprint arXiv:2404.07003.

[23] Gupta, A., Krishnan, U. M., Mandal, T. K., Chowdhury, R., & Nguyen, V. P. (2022). An adaptive mesh refinement algorithm for phase-field fracture models: Application to brittle, cohesive, and dynamic fracture. *Computer Methods in Applied Mechanics and*

*Engineering*, *399*, 115347.

[24] Xu, W., Li, Y., Li, H., Qiang, S., Zhang, C., & Zhang, C. (2022). Multi-level adaptive mesh refinement technique for phase-field method. *Engineering Fracture Mechanics*, *276*, 108891.

[25] Raheem, A. H. A., Mahdy, M., & Mashaly, A. A. (2019). Mechanical and fracture mechanics properties of ultra-high-performance concrete. *Construction and Building Materials*, *213*, 561-566.

[26] Xiong, Z., Liu, X., Di, D., Rittich, N., & Feldmann, M. (2024). Fatigue life prediction of orthotropic steel decks based on Phase Field model. In *IABSE Symp Manch* (pp. 328-335).

[27] Niu, Y., Huang, H., Wei, J., Jiao, C., & Miao, Q. (2022). Investigation of fatigue crack propagation behavior in steel fiber-reinforced ultra-high-performance concrete (UHPC) under cyclic flexural loading. *Composite Structures*, *282*, 115126.

[28] de Almeida Cerqueira, N. R., de Alencar Monteiro, V. M., de Souza, F. R., Cardoso, D. C. T., & de Andrade Silva, F. (2023). Mechanical degradation of ultra-high performance concrete under flexural fatigue loading. Construction and Building Materials, 391, 131877.

[29] Pise, M., Gebuhr, G., Brands, D., Schröder, J., & Anders, S. (2023). Phase-field modeling for failure behavior of reinforced ultra-high performance concrete at low cycle fatigue. *PAMM*, *23*(3), e202300233.

[30] Zhao, B., Geng, C., Song, Z., Pan, J., Chen, J., Xiao, P., ... & Yi, H. (2024). A fatigue fracture phase field model considering the effect of steel fibers in UHPC. *Engineering Fracture Mechanics*, *300*, 109981.

[31] Feldmann, M., & Schaffrath, S. (2018). Application of damage theory to structures made from high-strength steels: Prediction of component strength and ductility. *Steel Construction*, *11*(4), 257-263.

[32] Yan, J. B., Liew, J. R., Zhang, M. H., & Wang, J. Y. (2014). Mechanical properties of normal strength mild steel and high strength steel S690 in low temperature relevant to Arctic environment. *Materials & Design*, *61*, 150-159.

[33] Meissner, M., Bartscha, H., & Feldmanna, M. Toughness-dependent net-section resistance and ductility behaviour of high-strength structural steels S700 to S960. Nordic Steel Construction Conference 2024 (NSCC 2024) , Luleå, Sweden, 26-28 June 2024

[34] Tong, L., Niu, L., Jing, S., Ai, L., & Zhao, X. L. (2018). Low temperature impact toughness of high strength structural steel. *Thin-Walled Structures*, *132*, 410-420.

[35] Zhao, Y., Tong, X., Wei, X. H., Xu, S. S., Lan, S., Wang, X. L., & Zhang, Z. W. (2019).

Effects of microstructure on crack resistance and low-temperature toughness of ultra-low carbon high strength steel. *International Journal of Plasticity*, *116*, 203-215.

[36] Stranghöner, N., Lorenz, C., Feldmann, M., Citarelli, S., Bleck, W., Münstermann, S., & Brinnel, V. (2018). Sprödbruchverhalten hochfester Schrauben großer Abmessungen bei tiefen Temperaturen. *Stahlbau*, *87*(1), 17-29.

[37] Wu, J. Y., Yao, J. R., & Le, J. L. (2023). Phase-field modeling of stochastic fracture in heterogeneous quasi-brittle solids. *Computer Methods in Applied Mechanics and Engineering*, *416*, 116332.

[38] Baktheer, A., Martínez-Pañeda, E., & Aldakheel, F. (2024). Phase field cohesive zone modeling for fatigue crack propagation in quasi-brittle materials. *Computer Methods in Applied Mechanics and Engineering*, *422*, 116834.

[39] Bourdin, B., Francfort, G. A., & Marigo, J. J. (2000). Numerical experiments in revisited brittle fracture. *Journal of the Mechanics and Physics of Solids*, *48*(4), 797-826.

[40] Bourdin, B., Francfort, G. A., & Marigo, J. J. (2008). The variational approach to fracture. *Journal of elasticity*, *91*, 5-148.

[41] Wu, J. Y., & Cervera, M. (2015). On the equivalence between traction-and stress-based approaches for the modeling of localized failure in solids. *Journal of the Mechanics and Physics of Solids*, *82*, 137-163.

[42] Navidtehrani, Y., Betegon, C., & Martinez-Paneda, E. (2021). A unified Abaqus implementation of the phase field fracture method using only a user material subroutine. *Materials*, *14*(8), 1913.

[43] Alessi, R., & Ulloa, J. (2023). Endowing Griffith's fracture theory with the ability to describe fatigue cracks. *Engineering Fracture Mechanics*, *281*, 109048.

[44] Marigo, J. J. (1985). Modelling of brittle and fatigue damage for elastic material by growth of microvoids. *Engineering Fracture Mechanics*, *21*(4), 861-874.

[45] Khalil, Z., Elghazouli, A. Y., & Martinez-Paneda, E. (2022). A generalised phase field model for fatigue crack growth in elastic–plastic solids with an efficient monolithic solver. *Computer Methods in Applied Mechanics and Engineering*, *388*, 114286.

[46] Xiong, Z., Liu, X., Li, J., Baktheer, A., Wolters, K., & Feldmann, M. (2025). An accelerated phase-field model for high-cycle fatigue behaviour in quasi-brittle materials. *Theoretical and Applied Fracture Mechanics*, 105003.

[47] Navidtehrani, Y., & Martinez-Paneda, E. (2021). A simple yet general ABAQUS phase field fracture implementation using a UMAT subroutine. *Eng. Sci*, *6*, 100050.

[48] Gljušćić, M., Lanc, D., Franulović, M., & Žerovnik, A. (2023). Microstructural analysis

of the transverse and shear behavior of additively manufactured CFRP composite RVEs based on the phase-field fracture theory. Journal of composites science, 7(1), 38.
[49] Saqif, M. A., & El-Tawil, S. (2023). Characterization of the tension softening behavior of UHPC. *Construction and Building Materials*, *409*, 134063.
[50] Liu, R., Ding, R., Nie, X., & Fan, J. (2024). Experimental study on the mechanical properties of non-reinforced UHPC element under biaxial stress states. *Construction and Building Materials*, *456*, 1`265.
[51] Wu, J. Y., & Nguyen, V. P. (2018). A length scale insensitive phase-field damage model for brittle fracture. *Journal of the Mechanics and Physics of Solids*, *119*, 20-42.

## Nomenclature

| | |
|---|---|
| PFM | Phase Field Models |
| PF-CZM | Phase-Field Cohesive Zone Model |
| HSS | High Strength Steel |
| UHPC | Ultra-High Performance Concrete |
| UMAT | User Material |
| HETVAL | Heat Flux |
| SDVs | Solution Dependent State Variables |
| CMOD | Crack Mouth Opening Displacement |
| CNT | Center Notch Tension |
| CTS | Compact Tension Specimen |
| $\Omega$ | Domain of the solid |
| $\mathbb{R}^n$ | *n*-dimensional Euclidean space |
| $\partial\Omega$ | Outer boundary of the solid |
| $\mathbb{R}^{n-1}$ | *n*-1 dimensional Euclidean space |
| $\mathbf{n}$ | Outer normal vector of the solid |
| $\boldsymbol{b}^*$ | Body force |
| $\partial\Omega_u$ | Dirichlet boundary |
| $\partial\Omega_t$ | Neumann boundary |
| $\boldsymbol{u}^*$ | Displacement |
| $\boldsymbol{t}^*$ | Applied tractions |
| $S$ | Sharp crack |
| $\mathbf{n}_S$ | Normal vector of the sharp crack |
| $B$ | Crack band |
| $\partial B$ | Outer boundary of the crack band |
| $\mathbf{n}_B$ | Outer normal vector of the crack band |
| $\varepsilon$ | Strain tensor |
| $W$ | Work density function |
| $\gamma$ | Density function of the crack surface |
| $d$ | Phase Field parameter |
| $G_f$ | Fracture energy |
| $\nabla$ | Gradient operator |
| $b$ | Length scale regularizing the sharp crack |
| $c_0$ | Scaling factor |
| $\alpha(d)$ | Crack geometric function |
| $\xi$ | Coefficient of the crack geometric function |
| $\omega(d)$ | Energy degradation function |
| $\boldsymbol{\sigma}$ | Stress tensor |
| $w$ | Crack opening |
| $f_t$ | Failure / uniaxial tensile strength |
| $E_0$ | Young's modulus |
| $\nu$ | Poisson's ratio |
| $d^*$ | Maximum damage |
| $k_0$ | Initial slope of the softening curve |
| $w_c$ | Ultimate crack opening of the softening curve |
| $a_1$, $a_2$, $a_3$, $\beta_k$, $\beta_w$ | Parameters controlling the degradation function |
| $m$ | Exponent controlling the degradation function |
| $\psi$ | Strain energy density |
| $\psi_0$ | Initial elastic strain energy density |
| $\mathbb{C}$ | Elastic stiffness |
| $\mathrm{S}$ | Compliance |
| $\bar{\boldsymbol{\sigma}}$ | Effective stress tensor |
| $Y$ | Conjugate energy release rate |
| $\mathrm{Y}$ | Crack driving force |
| $\delta$ | Variation operator |
| $\Delta$ | Laplacian operator |
| $\bar{\boldsymbol{\sigma}}_{\mathrm{eq}}$ | Equivalent stress |
| $\rho_s$ | Ratio of uniaxial compressive strength to uniaxial tensile strength |
| $f_c$ | Uniaxial compressive strength |
| $\bar{\boldsymbol{\sigma}}_1$ | First principal stress |
| $\bar{I}_1$ | First invariant |
| $\bar{J}_2$ | Second invariant |
| $A_0$ | Coefficient of the generic stress-based failure criterion |
| $\sigma_1$, $\sigma_2$ | Biaxial stresses |
| $d_{t+\Delta t}$ | Phase Field parameter in the current time increment |
| $d_t$ | Phase Field parameter in the previous time increment |
| $\mathscr{H}$ | Historical field variable of the crack driving force |
| $\mathscr{H}_{\min}$ | Minimum value of the crack driving force |
| $\bar{\alpha}(t)$ | Cumulative historical variable |
| $t$ | Pseudo time |
| $f(\bar{\alpha}(t))$ | Fatigue degradation function |
| $\alpha_T$ | Fatigue damage threshold |
| $k_f$ | Fatigue damage accumulation parameter |
| $dN$ | Cyclic increment |
| $dD$ | Damage increment |
| $D$ | Damage variable |
| $D_c$ | Damage threshold |
| $L$ | Ratio of the average load to the maximum load |
| $c$ | Constant reflecting the average stress effect |
| $\hat{\sigma}_n$ | Maximum principal stress |
| $A_D$, $n_D$, $k_s$ | Parameters controlling the accelerated algorithm |
| $k$ | Thermal conductivity |
| $c_\rho$ | Specific heat capacity |
| $\rho$ | Density |
| $T$ | Temperature field |
| $r$ | Internal heat source |
| $\mathbf{K_u}$ | Stiffness matrix related to displacement |
| $\mathbf{K}_d$ | Stiffness matrix related to Phase Field parameter |
| $\mathbf{R_u}$ | Residual vector related to displacement |
| $\mathbf{R}_d$ | Residual vector related to Phase Field parameter |
| $k_1$, $k_2$ | Parameters controlling the softening curve |
| $S_{max}$ | Upper limit load |
| $P_{0.5mm}$ | Bearing capacity of the UHPC notched beam when the CMOD reaches 0.5mm |
| $N$ | Number of cycles |
| $N_f$ | Fatigue life |
| $S_{min}$ | Lower limit load |
| $P_{max}$ | Ultimate bearing capacity |
| $\theta$ | Loading angle |